\documentstyle[epsfig,psfig]{mn}     
     
\newif\ifAMStwofonts     
     
\ifoldfss     
  \ifCUPmtlplainloaded \else     
    \NewTextAlphabet{textbfit} {cmbxti10} {}     
    \NewTextAlphabet{textbfss} {cmssbx10} {}     
    \NewMathAlphabet{mathbfit} {cmbxti10} {} 
    \NewMathAlphabet{mathbfss} {cmssbx10} {} 
  \fi     
  \ifAMStwofonts     
    \ifCUPmtlplainloaded \else     
      \NewSymbolFont{upmath} {eurm10}     
      \NewSymbolFont{AMSa} {msam10}     
      \NewMathSymbol{\upi}     {0}{upmath}{19}     
      \NewMathSymbol{\umu}     {0}{upmath}{16}     
      \NewMathSymbol{\upartial}{0}{upmath}{40}     
      \NewMathSymbol{\leqslant}{3}{AMSa}{36}     
      \NewMathSymbol{\geqslant}{3}{AMSa}{3E}

       \let\le=\leqslant     
       \let\ge=\geqslant     
    \fi     
  \fi     
\fi 
     
\ifnfssone     
  \newmathalphabet{\mathit}     
  \addtoversion{normal}{\mathit}{cmr}{m}{it}     
  \addtoversion{bold}{\mathit}{cmr}{bx}{it}     
  \newmathalphabet{\mathbfit} 
  \addtoversion{normal}{\mathbfit}{cmr}{bx}{it}     
  \addtoversion{bold}{\mathbfit}{cmr}{bx}{it}     
  \newmathalphabet{\mathbfss} 
  \addtoversion{normal}{\mathbfss}{cmss}{bx}{n}     
  \addtoversion{bold}{\mathbfss}{cmss}{bx}{n}     
  \ifAMStwofonts     
    \ifCUPmtlplainloaded \else     
      \UseAMStwoboldmath     
      \makeatletter     
      \new@mathgroup\upmath@group     
      \define@mathgroup\mv@normal\upmath@group{eur}{m}{n}     
      \define@mathgroup\mv@bold\upmath@group{eur}{b}{n}     
      \edef\UPM{\hexnumber\upmath@group}     
      \new@mathgroup\amsa@group     
      \define@mathgroup\mv@normal\amsa@group{msa}{m}{n}     
      \define@mathgroup\mv@bold\amsa@group{msa}{m}{n}     
      \edef\AMSa{\hexnumber\amsa@group}     
      \makeatother     
      \mathchardef\upi="0\UPM19     
      \mathchardef\umu="0\UPM16     
      \mathchardef\upartial="0\UPM40     
      \mathchardef\leqslant="3\AMSa36     
      \mathchardef\geqslant="3\AMSa3E     

       \let\le=\leqslant     
       \let\ge=\geqslant     
    \fi     
  \fi     
\fi 
     
\ifnfsstwo     
  \DeclareMathAlphabet{\mathbfit}{OT1}{cmr}{bx}{it}     
  \SetMathAlphabet\mathbfit{bold}{OT1}{cmr}{bx}{it}     
  \DeclareMathAlphabet{\mathbfss}{OT1}{cmss}{bx}{n}     
  \SetMathAlphabet\mathbfss{bold}{OT1}{cmss}{bx}{n}     
  \ifAMStwofonts     
    \ifCUPmtlplainloaded \else     
      \DeclareSymbolFont{UPM}{U}{eur}{m}{n}     
      \SetSymbolFont{UPM}{bold}{U}{eur}{b}{n}     
      \DeclareSymbolFont{AMSa}{U}{msa}{m}{n}     
      \DeclareMathSymbol{\upi}{0}{UPM}{"19}     
      \DeclareMathSymbol{\umu}{0}{UPM}{"16}     
      \DeclareMathSymbol{\upartial}{0}{UPM}{"40}     
      \DeclareMathSymbol{\leqslant}{3}{AMSa}{"36}     
      \DeclareMathSymbol{\geqslant}{3}{AMSa}{"3E}     

       \let\le=\leqslant     
       \let\ge=\geqslant     
    \fi     
  \fi     
\fi 
     
\ifCUPmtlplainloaded \else     
  \ifAMStwofonts \else 
    \def\upi{\pi}     
    \def\umu{\mu}     
    \def\upartial{\partial}     
  \fi     
\fi

\title{A New approach for a Galactic Synchrotron Polarized Emission Template in     
	the Microwave Range}     
\author[G. Bernardi et al.]     
       {G. Bernardi,$^{1,2}$ E.~Carretti,$^1$ R.~Fabbri,$^3$ C.~Sbarra,$^1$     
        S.~Poppi,$^4$ S.~Cortiglioni$^1$\\     
        $^1$C.N.R./I.A.S.F. Bologna, Via Gobetti 101,      
	         I-40129 Bologna, Italy\\     
        $^2$Dipartimento di Astronomia, Universit\`a degli Studi di Bologna,      
           Via Ranzani 1, I-40127 Bologna, Italy\\     
        $^3$Dipartimento di Fisica, Universit\`a di Firenze, Via Sansone 1,     
            I-50019 Sesto Fiorentino (FI), Italy\\     
        $^4$C.N.R./I.R.A. Bologna, Via Gobetti 101,      
	         I-40129 Bologna, Italy\\     
 	}     
\date{24 October 2002}     
     
\pagerange{\pageref{firstpage}--\pageref{lastpage}}     
\pubyear{0000}     
     
\begin{document}     
     
\maketitle     
     
\label{firstpage}     
     
\begin{abstract}     
We present a new approach in modelling the polarized Galactic synchrotron 
emission in the microwave range (20-100~GHz), where this radiation is expected  
to play the leading role in contaminating the Cosmic Microwave Background (CMB)
data. Our method is based on real surveys and aims at providing the real
spatial distributions of both polarized intensity and polarization angles. Its
main features are the modelling of a polarization horizon to determine the
polarized intensity and the use of starlight optical data to model the
polarization angle pattern. Our results are consistent with several existing
data, and our template is virtually free from Faraday rotation effects as
required at frequencies in the cosmological window. 
\end{abstract}     
     
     
 \section{Introduction} \label{intro} 
       
The polarized component of the diffuse background emission in the microwave
range is of great interest for both the Galactic structure and the CMB.
Actually, its measurement leads to probing the structure of the interstellar    
medium (ISM) and the Galactic magnetic field. Moreover, the detection of CMB
Polarization (CMBP) allows the investigation of the early Universe.     
     
CMB anisotropies and polarization are powerful tools to determine cosmological 
parameters (Sazhin \& Benitez 1995, Zaldarriaga, Spergel \& Seljak 1997, 
Kamionkowski \& Kosowsky 1998). However, although anisotropies have been      
already detected  and space missions      
(MAP\footnote{http://map.gsfc.nasa.gov/}, PLANCK\footnote{http://astro.estec.esa.nl/SA-general/Projects/Planck/})   
are expected to make all-sky surveys down to $0^\circ.1$ angular resolution,   
CMBP still represents a challenge for astronomers. The first    
detection has been just claimed by DASI (Kovac et al. 2002) and         
several experiments will address it soon 
(SPOrt\footnote{http://sport.bo.iasf.cnr.it} (see Carretti et al. 2002, 
Cortiglioni et al. 2002), MAP, PLANCK, B2K2 (Masi et al., 2002), BaR-SPOrt 
(Zannoni et al., 2002) and AMiBA (Kesteven et al., 2002) among the others).    
     
Besides the CMBP low emission level      
(3-4~$\mu$K on sub-degree scales and      
$< 1\, \mu$K on large ones), difficulties in its detection are mainly     
related to the presence of foreground noise from Galactic and   
extragalactic sources.     
Extragalactic foregrounds essentially consist of radio and    
infrared discrete     
sources, whereas Galactic foregrounds are generated by synchrotron,      
free-free, thermal dust and spinning/magnetic dust emissions.   
     
Synchrotron polarized emission should represent the most relevant 
foreground in  the microwave range:      
free-free is fainter ($< 4\, \mu$K at 30~GHz in total intensity, see      
Reynolds \& Haffner 2000) and almost unpolarized, whereas thermal dust has a     
polarization degree much smaller than synchrotron (Prunet et al. 1998,      
Tegmark et al. 2000).      
Evidence for spinning or magnetic dust emission has been found      
(Kogut et al. 1996, de Oliveira--Costa et al., 2002) but it seems to play an     
important role only up to $\sim 50$~GHz. Moreover, it should    
have a low polarization degree (Lazarian \& Prunet 2002).     
     
In spite of its     
importance, synchrotron emission is scarcely surveyed: existing data mainly 
cover the Galactic Plane area at frequencies up to 2.7~GHz, far away from the 
cosmological window (Duncan et al. 1997, hereafter D97, Duncan et al. 1999,   
hereafter D99, Uyaniker et al. 1999, Gaensler et al. 2001, Landecker et al.   
2002). The Leiden data (Brouw \& Spoelstra 1976, hereafter BS76) cover high 
Galactic latitudes, but are limited to  $<$~1.4~GHz and are largely 
undersampled. This situation makes having a reliable synchrotron  
polarized emission template in the 20-100~GHz range very important.    
This would allow, for instance, reliable numerical     
simulations to set-up and test destriping techniques or foreground separation     
methods (Revenu et al. 2000, Sbarra et al. 2003, Tegmark et al. 2000 and     
references therein).    
At present, only toy models exist, which do not account for the real   
spatial distribution     
of both polarization intensity and polarization angles    
(Kogut \& Hinshaw 2000,   Giardino et al. 2002).     
     
In this paper we present a new approach in modelling the Galactic diffuse 
synchrotron polarized emission in the 20-100~GHz range. It is based on real 
surveys and fitted to the real spatial distribution of both polarized intensity 
and polarization angles. Low frequency data are used to model the polarized 
intensity and optical starlight is used to model polarization angles. This 
allows the construction of $Q$ and $U$ maps covering about half of the sky 
with the SPOrt experiment angular resolution (FWHM~$= 7^\circ$). Although our 
work is finalized to the SPOrt experiment, the method is general enough to be 
suitable also for smaller angular scales as soon as complete sets of data with 
sub-degree angular resolution will be available.    
     
The great advantage of this new approach    
is to produce $Q$ and $U$ maps free from Faraday rotation, allowing     
a direct extrapolation to the cosmological window.      
     
The outline of the paper is as follows:      
the synchrotron polarized emission model and the procedure to build    
the template are presented in Section~\ref{model},    
results and comparisons with existing data are described in
Section~\ref{results} and~\ref{test}, respectively, whereas Section~\ref{conc}
contains the conclusions.      
     
 \section{THE MODEL} \label{model}

 \subsection{Ingredients}\label{ingredients}    
      
 The aim of our work is to generate template maps of the two linear Stokes      
 parameters $Q$ and $U$ of the Galactic synchrotron polarized radiation in the     
 cosmological window near 100~GHz with the SPOrt angular resolution    
(FWHM $=7^\circ$). We divide the problem in two parts:    
\begin{enumerate}    
 \item{} constructing a polarized intensity ($I_p$) map:     
        it can be obtained from existing total intensity ($I$) sky surveys    
        assuming a model linking the polarized to      
        the total intensity synchrotron emission;     
 \item{} building a map of polarization angles    
        not affected by the effects of Faraday rotation. At present,    
        only optical starlight data fulfil this requirement.    
\end{enumerate}     
The Haslam map (Haslam et al. 1982) is the most complete sky survey at    
radio wavelenghts, where synchrotron emission is dominant. It is a full-sky map
at $408$~MHz with a resolution of $51$~arcmin obtained combining observations
taken with different radiotelescopes. However, it is not perfect for our aims
since the free-free emission is still significant, expecially in the Galactic
plane (Reich \& Reich 1988, hereafter RR88). Consequently, identification and
subtraction of this contribution is mandatory. As described in
Section~\ref{dodelson}, we perform this separation using the Dodelson (1997)
formalism, which requires a second map at different frequency. We use the Reich
(1982) map at 1.4~GHz with an angular resolution of $35$~arcmin, the only other
available survey with absolute calibration covering a large part of the sky
($\delta \ge -17^\circ$).    
    
The polarization angles are taken from the Heiles starlight polarization     
catalogue (Heiles 2000). These data have many advantages with respect to those 
in the radio band: they are free from Faraday rotation and they cover almost all 
the sky. The catalogue lists polarization degree, position angle and distance    
of about $9000$~stars from both hemispheres.   
Further properties of this catalogue and considerations    
confirming its validity as a polarization angle pattern for our    
model are described in section~\ref{starlight}.

 \subsection{Synchrotron Intensity Map}     
 \label{dodelson}     
   
 The different frequency behaviour of free-free and synchrotron emissions    
 allows the application of the Dodelson technique (Dodelson 1997) to    
 Haslam (0.408~GHz) and Reich (1.4~GHz) maps.    
 The original Dodelson formalism is centred on the CMB frequency    
 dependence, so that here a slight modification is introduced to adjust    
 the method to the synchrotron--free-free case.    
 For each pixel $i$ a vector ${\bf T}_i$ is defined whose elements    
 are the pixel antenna temperatures at the two frequencies. Its expression in 
 terms of synchrotron and free-free components is given by:     
 \begin{equation}      
 {\bf T}_i = {\bf{T}}^s_i + {\bf{T}}^{f \! f}_i +  {\bf N}_i,      
 \end{equation}    
 where  ${\bf{T}}^s_i$ and $ {\bf{T}}^{f \! f}_i$ are the synchrotron    
 and free-free contributions, respectively, and ${\bf N}_i$ is the    
 noise.    
       
 Assuming the noise from the two maps is completely uncorrelated     
 (it comes from different experiments), the correlation matrix       
 \begin{equation}      
 C_{ab,i} = \left< N_{a,i} N_{b,i} \right>      
 \end{equation}      
 is diagonal with pixel variances as elements ($a$ and $b$ indicate    
 the frequencies).    
     
 Provided the frequency behaviours (shapes) ${\bf{F}}^s_i$ and ${\bf{F}}^{f \! 
 f}_i$ are known, the estimators ${\bf{\Theta}}^s_i$ and ${\bf{\Theta}}^{f \! 
 f}_i$ of ${\bf{T}}^s_i$ and ${\bf{T}}^{f \! f}_i$ can be expressed as: 
 \begin{eqnarray}      
 {\bf\Theta}^s_i        &=&  \theta^s_i {\bf {F}}^s_i \\      
 {\bf{\Theta}}^{f \! f}_i &=&  \theta^{f \! f}_i {\bf{F}}^{f \! f}_i       
 \end{eqnarray}      
 where $\theta^s_i$ and $\theta^{f \! f}_i$ are the unknown amplitudes of    
 synchrotron and free-free, respectively.    
    
 In the range $0.408-1.4$~GHz free-free and synchrotron are known to   
 follow power laws, so that their shapes are    
 \begin{eqnarray}      
 {F}^s_{i,a}        &=& \nu_a^{-\beta_i}  \\    
 {F}^{f \! f}_{i,a} &=& \nu_a^{-\alpha}    
 \end{eqnarray}    
 with the free-free spectral index $\alpha = 2.1$ (RR88).   
 The frequency behaviour of synchrotron radiation depends on the energy      
 distribution of relativistic electrons and is spatially varying       
 across the sky.   
      
 We define a scalar product between vectors as     
 \begin{equation}      
 {\bf R} \cdot {\bf S} = \sigma^{(0)^2}_\theta\sum^{2}_{a,b=1}R_aC^{-1}_{ab}S_b. 
 \end{equation}    
 This expression is similar to that of Dodelson,   
 apart from the normalization factor         
 \begin{equation}      
 \sigma^{(0)^2}_\theta =      
 \frac{1}{\sum^{2}_{a,b=1}F^{f \! f}_aF_b^{f \! f}C^{-1}_{ab}}, 
 \end{equation}     
 which here is based on the free-free shape rather than on that of CMB.     
 Again following Dodelson, the best estimate    
 of the amplitudes is given by    
 \begin{eqnarray}     
 \theta^s_i        &=& \sum_{j}K^{-1}_{1j} {\bf {F}}^j_i\cdot {\bf T}_i ,\\     
 \theta^{f \! f}_i &=& \sum_{j}K^{-1}_{0j} {\bf {F}}^j_i\cdot {\bf T}_i 
 \hspace{1cm} \mbox{with } j = s,f \! f,    
 \label{eq10}    
 \end{eqnarray}     
 where the matrix {\bf K} is defined as     
 \begin{equation}      
 K_{ij} = {\bf F}^i \cdot {\bf F}^j.      
 \end{equation}    
 As a result, the two components are separated providing the two maps    
 of synchrotron ($\theta^s_i$) and free-free ($\theta^{f \! f}_i$).    
 
 The spectral index $\beta_i$ has been modelled using the analysis of RR88    
 who compared data at $408$ MHz and $1.4$~GHz obtaining the following results:    
\begin{itemize}    
 \item{} $\beta \sim 2.85$ toward the Galactic    
   anticentre at $b = 0^\circ$ of Galactic latitude, with a flattening with    
   incresing $z$ ($z$ is the height above the Galactic plane);   
 \item{} $\beta \sim 3.1$ in the inner disk (Galactocentric distance    
     $<8$~kpc and $|z| < 2$~kpc). In this region the spectral index is nearly    
     constant, beginning to decrease from $|b| \simeq 10^\circ$ until reaching    
     the value of $\beta \sim 2.65$ at $|b| \simeq 30^\circ$;   
 \item{} $\beta\sim 2.65$ for $|b| > 30^\circ$ independently of the    
        Galactic longitude $l$.   
\end{itemize}    
  The transition from the inner disk region to the rest of the Galactic plane    
 occurs between $l = 45^\circ$ and $l = 55^\circ$, where a flattening    
 from $\beta = 3.1$ to $\beta = 2.85$ is observed. From spectral index profiles 
 (cfr. Figure~5 in RR88) we find the linear behaviour to be a good approximation 
 for $\beta$ in the transition regions. From these considerations,    
 we model the distribution of synchrotron spectral indeces as follows    
 (see Figure~\ref{spectral_index}):    
 \begin{figure}      
 \begin{center}      
 \includegraphics[width=0.6\hsize,angle=90]{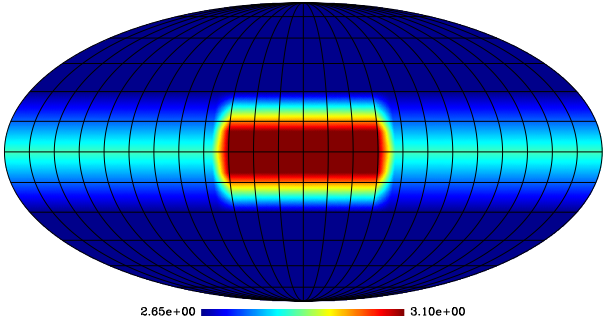}      
 \end{center}     
 \caption{Synchrotron spectral index map in the 0.408--1.4~GHz range   
          (see text and RR88 for details). The map is in Galactic coordinates 
	  with the Galactic Centre in the middle and Galactic longitudes 
	  increasing toward left.}       
 \label{spectral_index}     
 \end{figure}      
 \begin{description}     
    \item{1)} a region towards the Galactic centre ($|b| < 10^\circ$ and $l    
         < 45^\circ$, $l > 315^\circ$) with $\beta = 3.1$;    
    \item{2)} a region towards the Galactic anticentre on the Galactic plane       
         ($b = 0^\circ$ and  $55^\circ < l < 305^\circ$) with $\beta = 2.85$;          
    \item{3)} a region at high Galactic latitude ($|b| > 30^\circ$) with       
         $\beta = 2.65$.    
    \item{4)} the spectral index follows a linear behaviour in the transition
    	regions and continuity is imposed at borders.
	Figure~\ref{spectral_behaviour} shows the spectral index behaviour in
	three special cases.
 \end{description}     
\begin{figure}      
\begin{center}      
\includegraphics[width=0.65\hsize,angle=90]{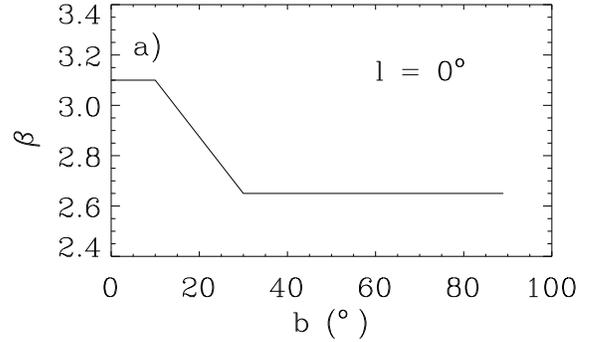}      
\includegraphics[width=0.65\hsize,angle=90]{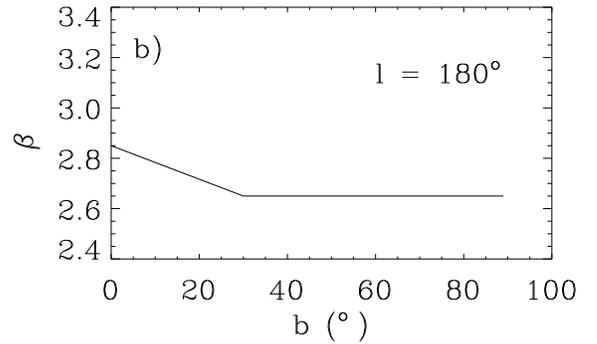}      
\includegraphics[width=0.65\hsize,angle=90]{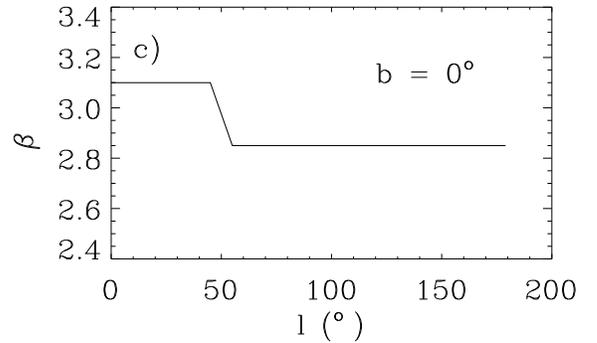}      
\end{center}     
\caption{Spatial behaviour of the synchrotron spectral index $\beta$   
         for three special    
         cases: $l = 0^\circ$ (a), $l=180^\circ$ (b)     
         and $b = 0^\circ$ (c).    
 }      
 \label{spectral_behaviour}     
 \end{figure}    
       
 \subsection{From Total to Polarized Intensity Map}    
\label{polHorSec}    
     
\begin{figure}      
 \begin{center}      
 \includegraphics[width=0.7\hsize]{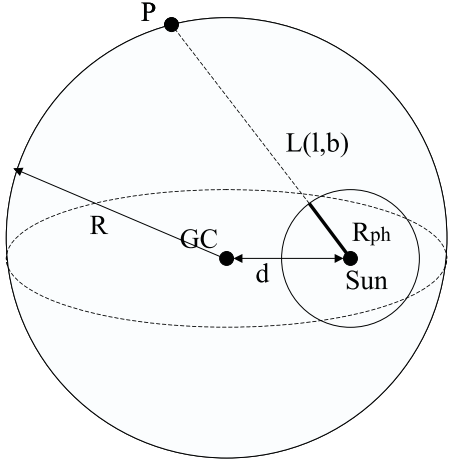}      
 \end{center}     
 \caption{Contributions to the Galactic polarized radiation are limited by a
 	polarization horizon, a local bubble of radius $R_{ph}$ centred on    
          the Sun (see text for details). The thickness $L(l,b)$ of the 
	  synchrotron emitting region of our simple spherical model (the 
	  distance of the halo point $P$ from the Sun position) is also shown.}      
 \label{fig:pluto}     
 \end{figure}

A common result of radio-surveys in polarization is the identification of    
two main components: a strong emission from discrete sources (Supernova Remnants 
- SNRs - and several sources with no $I$-counterpart) and a weaker, diffuse 
emission from a background component (D97, D99, Landecker et al. 2002, Gaensler 
et al. 2001) that appears to be rather constant with the longitude independently 
of both angular resolution and frequency. This isotropic background suggests 
the presence of a polarization horizon, a local screen beyond which the 
polarized emission is cancelled out (D97, Gaensler at al. 2001, Landecker et al. 
2002). The horizon can be imagined as a sort of bubble centred in the observer 
position (see Figure~\ref{fig:pluto}): the net polarized signal is only that 
integrated along the line of sight out to the horizon, whereas signals beyond 
the horizon are depolarized by variations of polarization angles (changes and 
turbulence in the Galactic magnetic field).    
   
 A further element suggesting   
  the existence of such a horizon comes from     
 Landecker et al. (2002), who   
 show that only the closest SNRs are well visible also in polarized emission,   
 whereas the most distant ones completely disappear.    
    
 The size of the horizon is not yet known: it depends on several effects along 
 the line of sight, like Galactic magnetic field turbulence and electron density 
 variations. However, it has been suggested that it can range from $2$~kpc 
 (Gaensler et al. 2001) up to 7~kpc (D97, Landecker et al. 2002), so that a few 
 kpc appeares to be a quite acceptable estimate.      
    
 The polarization horizon allows us to model the relation between    
 polarized and  total intensity synchrotron emissions.     
 Given the mean total synchrotron emissivity $J^s(\nu,l,b)$    
 at Galactic coordinates $(l,b)$ and the thickness $L(l,b)$ of the synchrotron 
 emitting region (see Figure \ref{fig:pluto}) in the same direction, the 
 brightness temperature      
 $T^s( \nu,l,b)$ at  frequency $\nu$ is:      
 \begin{equation} \label{eq4}     
 T^s( \nu,l,b) =  \frac{c^2}{ 2 K \nu^2} J^s( \nu,l,b) \, L(l,b),    
 \end{equation}       
 where $K$ is the Boltzmann constant.     
The thickness $L(l,b)$ depends on the geometrical model describing the space 
distribution of the relativistic--electron gas responsible for synchrotron 
emission. As a first step in modelling the polarized  synchrotron radiation     
we consider the simplest case where the gas is uniformly    
distributed in the Galactic halo. This is represented by    
a sphere of radius $R = 15$~kpc centred into the Galactic Centre (GC).     
Thus, in our simple case the line of sight $L(l,b)$ is    
the distance between the Sun and the edge of this sphere:    
        
\begin{equation} \label{eq5}     
  L(l,b) = d \, \cos(b) \, \cos(l) \, \left[ 1 + \sqrt{1 + \frac{(R^2/d^2 -      
  1)}{\cos^2(b) \, \cos^2(l)}} \right],   
\end{equation}      
where $d=9$~kpc is the Sun distance from GC (See Figure~\ref{fig:pluto} and     
Appendix~\ref{appLoS} for details).    
    
The polarized brightness temperature $T_p^s$ can be    
similarly defined provided   
the emission is integrated out to the polarization horizon $R_{ph}$ and    
a polarization degree $p$ is introduced:       
\begin{equation} \label{eq6}     
T^s_p( \nu,l,b) = \frac{c^2}{ 2 K \nu^2} \,p\,J^s( \nu,l,b) \, R_{ph}.      
\end{equation}       
Finally, equations~(\ref{eq4})~and~(\ref{eq6}) provide     
the relation between polarized and total intesity emissions:   
\begin{equation} \label{eq8}     
        T^s_p(\nu,l,b) = p \frac{R_{ph}}{L(l,b)} T^s( \nu,l,b)    
\end{equation}    
The quantity $p \, R_{ph}$ is unknown and represents a free parameter   
to be calibrated with real data.    
    
 \subsection{Polarization angle map}\label{starlight}     
    
 The propagation of an electromagnetic wave of wavelength $\lambda$ through a      
 plasma in presence of a magnetic field ${\bf B}$ is affected by Faraday 
 rotation. The net effect is a change in the polarization angle $\phi$ by     
 \begin{eqnarray} \label{eq11}     
 \Delta \phi &=& \mbox{RM} \, \lambda^2 \\    
 \mbox{RM} &=& 812 \int n_e (\mbox{cm}^{-3})\,      
 {\bf B} (\mu \mbox{G})\cdot d{\bf l} ({\rm kpc}) \, \mbox{rad m}^2 \nonumber    
 \end{eqnarray}	      
 where RM is the rotation measure, $n_e$ is the plasma electron density and 
 $d{\bf l}$ is the infinitesimal path along the line of sight.    
     
 Estimates of RM from extragalactic radio sources give typical values  ranging 
 from tens to hundreds rad/m$^2$ at medium and high Galactic latitudes, and from 
 tens to thousands rad/m$^2$ in the Galactic plane (Simard-Normandin \& Kronberg 
 1980, Sofue \& Fujimoto 1983, hereafter SF83, Brown \& Taylor 2001, hereafter 
 BT01). In particular, the behaviour along the Galactic plane is well fitted by 
 (BT01)    
 \begin{eqnarray} \label{eq12}     
 \mbox{RM}(l) &=& \mbox{RM}_0\cos(l - l_0) \\     
 \mbox{RM}_0 &=& -183 \pm 14 \mbox{ rad/m}^2 \nonumber\\    
 l_0 &=& 84^\circ \pm 4^\circ. \nonumber     
 \end{eqnarray}      
 Brown \& Taylor suggest that the modulation in RM occurs because {of} a local 
 constant magnetic field.      
   
 Equation~(\ref{eq12}) gives, in the frequency range of the cosmological window 
 ($20-100$~GHz), negligible Faraday rotation effects ($\pm 2.5^\circ$ at 
 20~GHz): all we need is a template of intrinsic polarization angles.   
 When used at $2.7$~GHz, which is the highest frequency of present polarization  
 surveys, equation~(\ref{eq12}) results in angular rotations up to $\pm 
 160^\circ$. This means that radio polarization data cannot    
 be used to build a reliable template of intrinsic polarization angles.    
    
 To overcome this problem we use the Heiles catalogue on starlight polarization, 
 the optical frequency being unaffected by Faraday rotation.    
   
 The polarization vector of  starlight is parallel to the Galactic magnetic 
 field $\bf B$ because of selective absorption by interstellar dust grains,  
 whose minor axis is aligned with $\bf B$ (Fosalba et al., 2001). Since the 
 synchrotron polarization vector is perpendicular to $\bf B$, starlight 
 polarization angles can be used as a template provided a $90^\circ$ rotation is 
 performed.     
      
\begin{figure}    
 \begin{center}      
 \includegraphics[width=1\hsize,angle=0]{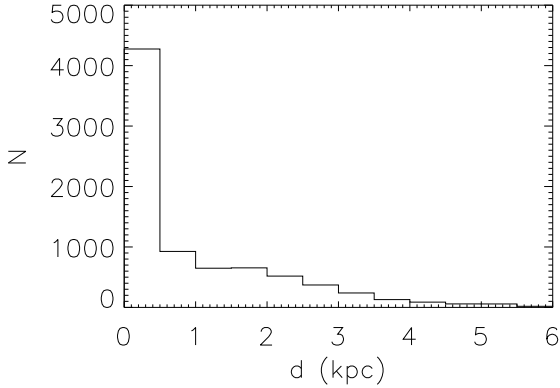}      
 \caption{Number of Heiles stars versus Sun distance.}      
 \label{distanze}   
 \end{center}      
 \end{figure}

 Most of the Heiles catalogue stars ($\simeq 87\%$) are within  $2$~kpc (see 
 Figure~\ref{distanze}), tracking the local magnetic field. Their 
 distance is in the order of the polarization horizon size (see Section 
 \ref{polHorSec}) confirming that the Heiles catalogue can be safely    
 used as a template for the polarization angles of synchrotron emission. The 
 main problems to face with when using Heiles polarization angles are the 
 irregular distribution of data and the variable sampling distance. However, 
 if the uniformity scale of the angles is compatible with the sampling distance 
 of the catalogue, the lack of data can be filled by linear interpolation.    
   
 An estimate of the uniformity scale can be obtained from Figures~9 and~10   
 of D97, showing that background emission regions have a polarization    
 angle pattern  varying  slowly on  scales of $5^\circ-10^\circ$.    
 Only areas with strong sources show a more complex structure, but their 
 modellization is out of the purposes of this paper.  
     
 The sampling distance of the Heiles catalogue (Figure~\ref{campionata})    
 is compatible with the uniformity scale everywhere but in the region centred 
 in ($l = 135^\circ, b = 40^\circ$) where it is greater than $10^\circ$. We 
 exclude this region from the interpolation procedure as well as from final 
 template maps.     
 
 We perform the interpolation by generating $Q$,~$U$ pairs corresponding to the 
 Heiles polarization angles $\theta$:    
 \begin{eqnarray}   
    Q_\theta &=& \cos\,(2\theta);\\   
    U_\theta &=& \sin\,(2\theta).   
 \end{eqnarray}   
 Then, for each pixel of the template map under construction we linearly 
 interpolate the  $Q_\theta$ and $U_\theta$  values of the three closest stars 
 and compute the corresponding polarization angle. The interpolation methods 
 uses parallel transport as described in Bruscoli et al. (2002). 
 \begin{figure}        
 \begin{center}      
 \includegraphics[width=0.6\hsize,angle=90]{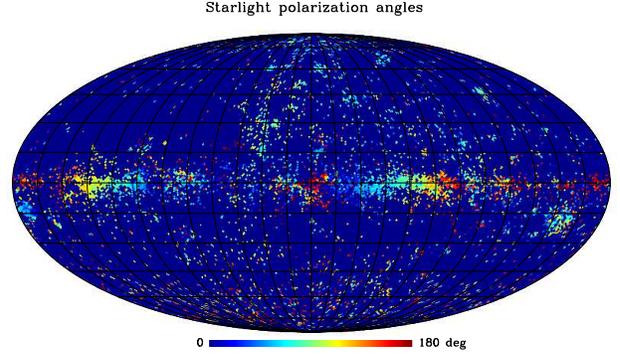}     
 \caption{Polarization angles of stars in the Heiles catalogue (degrees):  
           each star  
           is represented by a pixel in a HEALPix format map with   
           $Nside = 128$ (pixel size $\simeq 0^\circ.5$)).  
           The map is in Galactic coordinates.}      
 \label{map_stars}     
 \end{center}  
 \end{figure}		      
 \begin{figure}        
 \begin{center}  
 \includegraphics[width=0.6\hsize,angle=90]{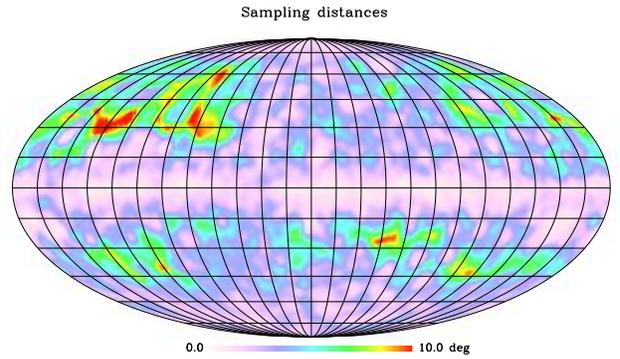}     
 \caption{Map of sampling distances (degrees) of the Heiles catalogue.   
          The map is in Galactic coordinates.}      
 \label{campionata}     
 \end{center}  
 \end{figure}

 \subsection{The Procedure} \label{proc}    
      
 The polarized synchrotron emission template is built as follows:    
 \begin{enumerate}    
 \item{} the CMB emission and the absolute calibration error are removed from 
	 both the Reich and the Haslam maps using values suggested in RR88 
	 ($3.7$~K and $2.8$~K for Haslam and Reich maps, respectively);   
 \item{} the resulting maps are resampled in HEALPix{\footnote
 	 {http://www.eso.org/science/healpix/}} format    
         with $N_{side} = 128$, corresponding to a pixel size   
         of about half a degree;   
 \item{} Reich map data are smoothed to the Haslam resolution     
        (FWHM~$ = 51$~arcmin);    
 \item{} the technique for component separation described in 
 	 Section~\ref{dodelson} is applied by using the synchrotron spectral 
	 index pattern previously described. This results in    
         the $I$ synchrotron emission map;    
 \item{} the relation between $I$ and $I_p$ described by equation~(\ref{eq8}), 
 	 which  
 	 introduces the effects of the polarization horizon, is used to provide 
	 the shape of the $I_p$ map, the parameter $p \, R_{ph}$ being still 
	 free. Its calibration is performed a posteriori on BS76 data and a 
	 detailed discussion is presented in Section~\ref{results};    
 \item{} from the $I_p$ map and from the angle map obtained from starlight data, 
	 $Q$ and $U$ maps are computed;   
 \item{} the $Q$ and $U$ maps produced in this way are convolved with a FWHM~$= 
 	 7^\circ$ Gaussian filter to match the SPOrt angular resolution. The 
	 smoothing procedure applies the parallel transport method described in 
	 Bruscoli et al. (2002) 
 \end{enumerate}    
      
 \section{The Polarized Synchrotron Template}\label{results}    
       
\begin{figure*}      
 \begin{center}  
 \includegraphics[width=0.3\hsize,angle=90]{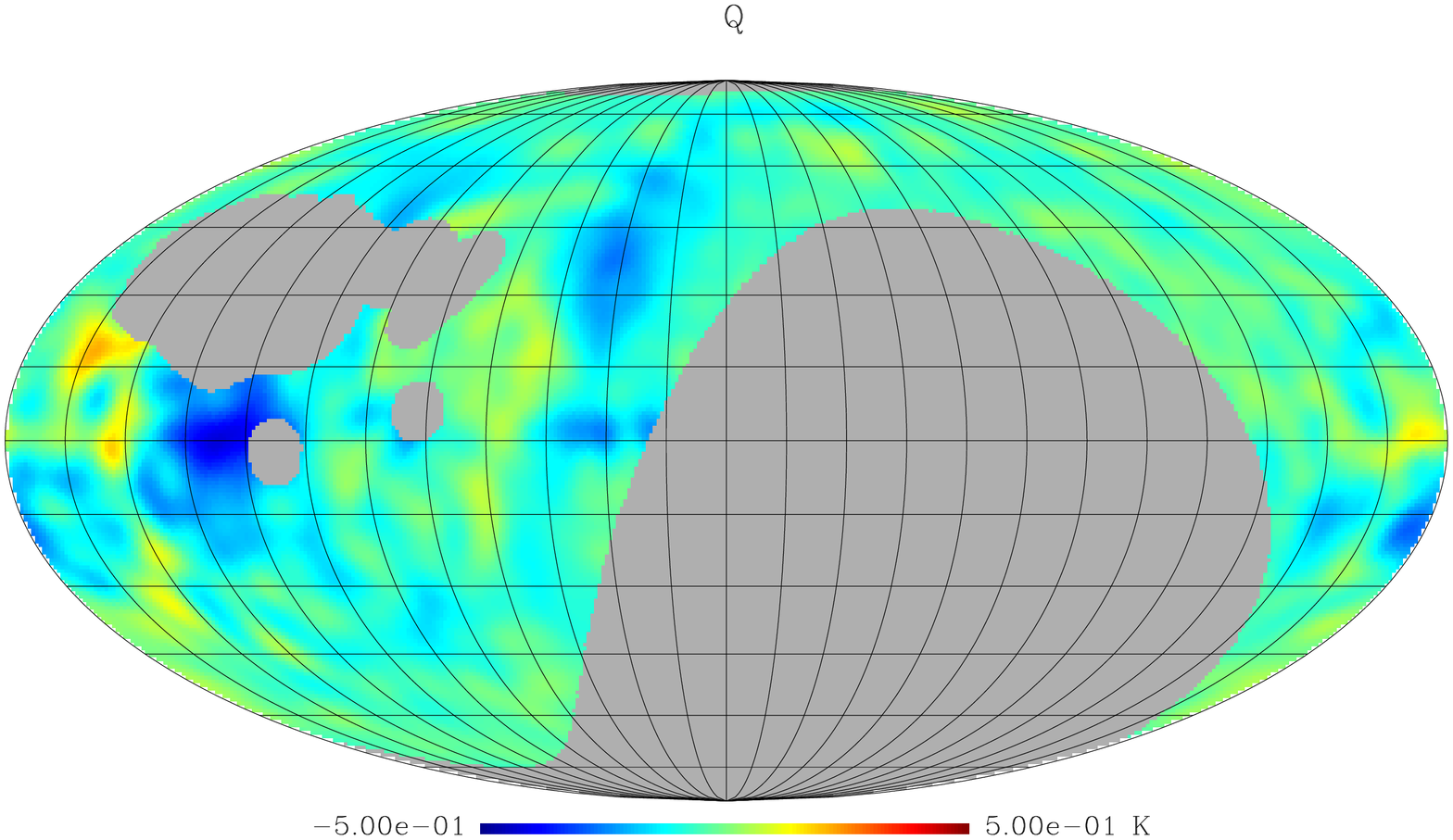}        
 \includegraphics[width=0.3\hsize,angle=90]{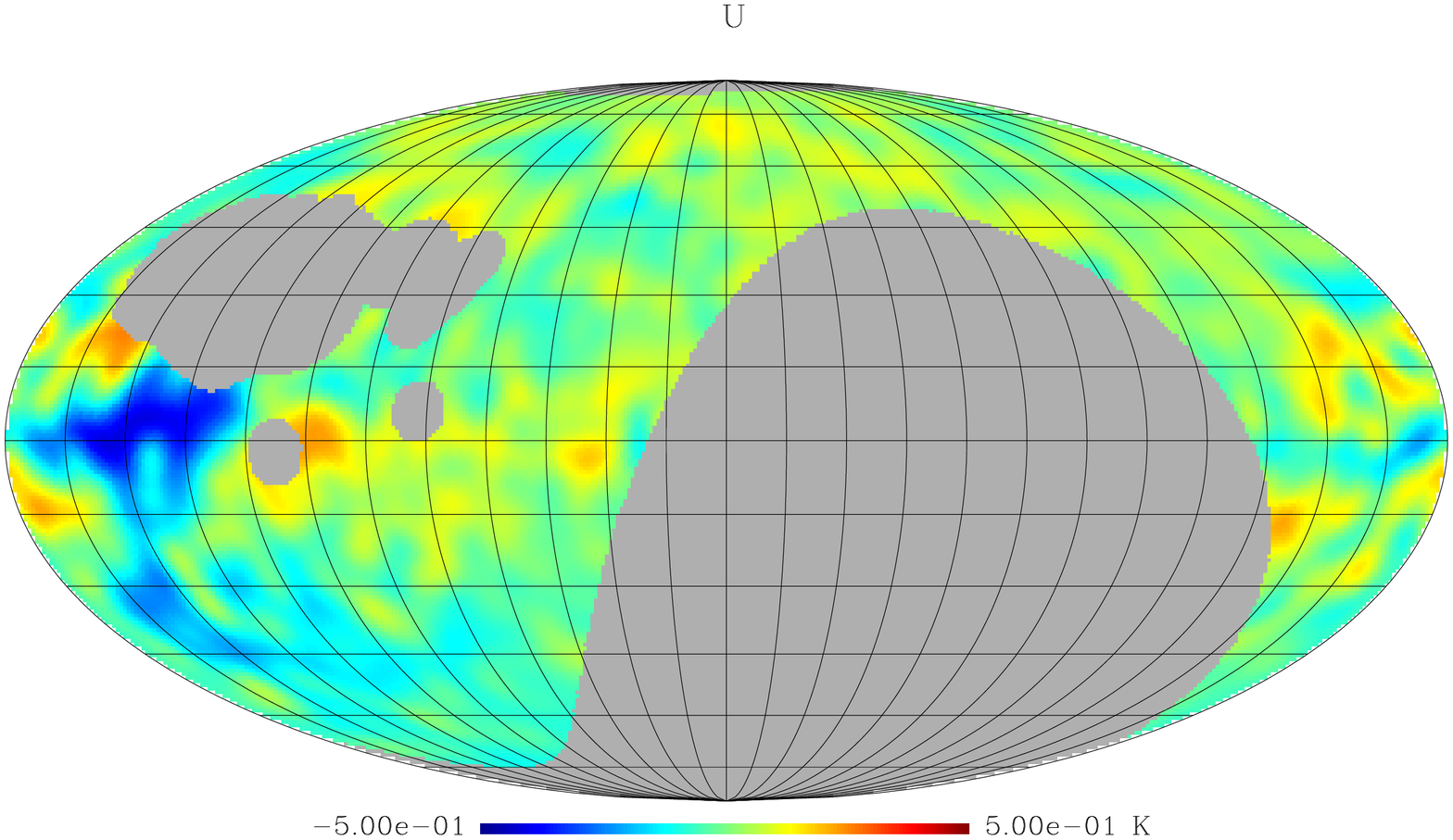}      
 \caption{$Q$ (left) and $U$ (right) maps of our synchrotron polarized   
 emission template (Kelvin). The maps are at 1.4~GHz convolved with    
 a FWHM$= 7^\circ$ Gaussian filter.}      
 \label{qu_maps}     
 \end{center}  
 \end{figure*}

\begin{figure*}      
 \begin{center}  
 \includegraphics[width=0.3\hsize,angle=90]{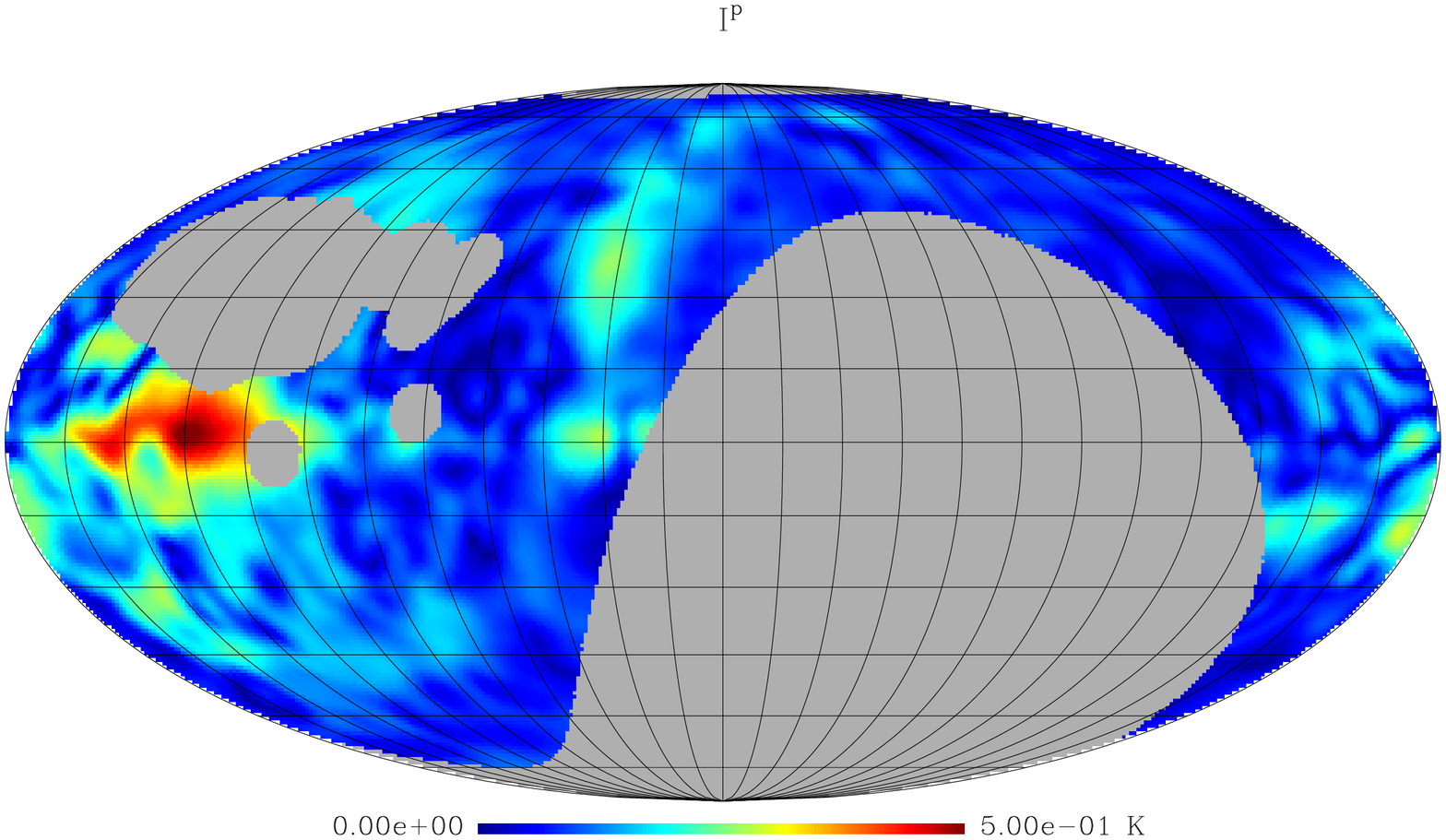}      
 \includegraphics[width=0.3\hsize,angle=90]{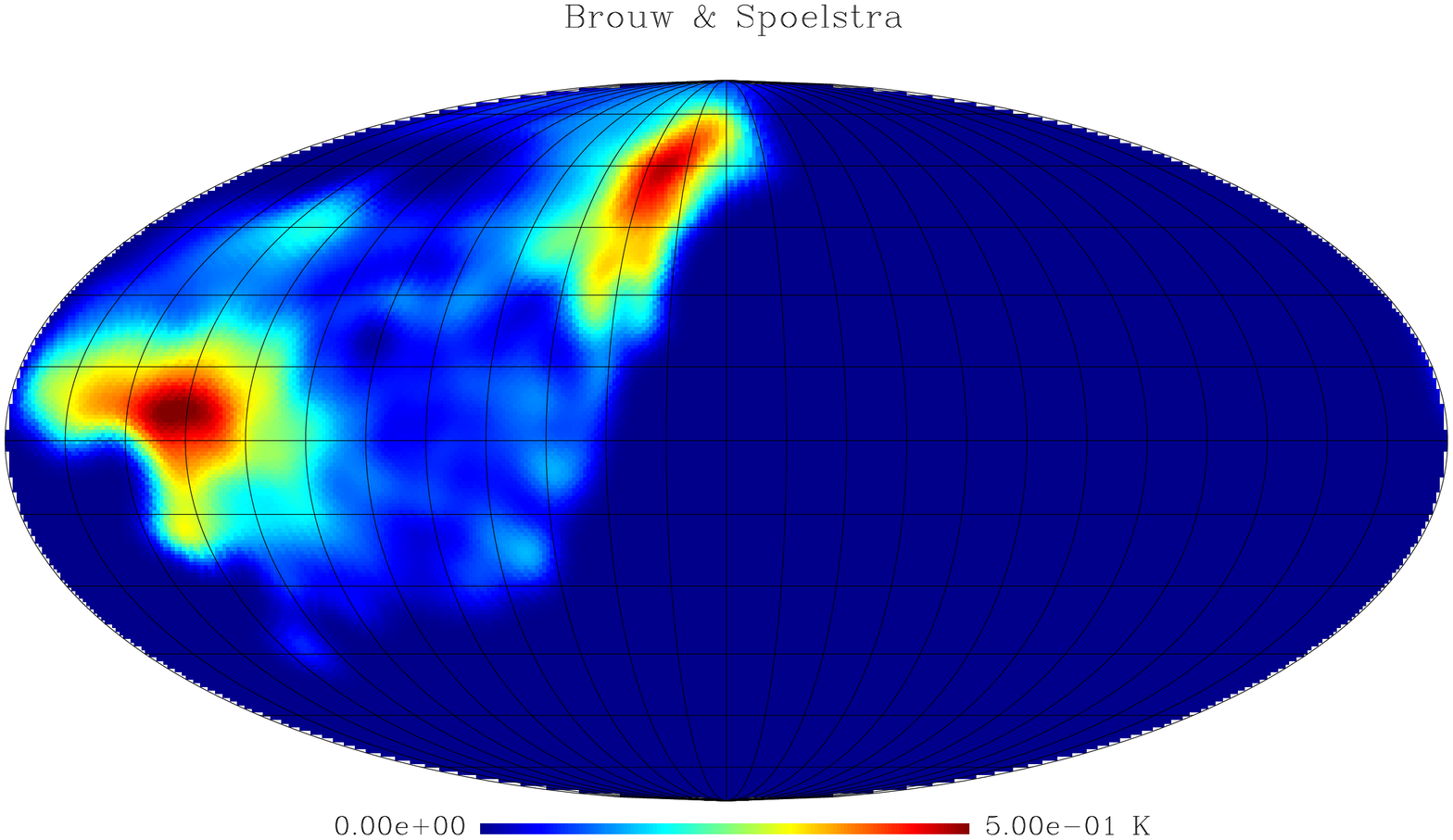}      
 \caption{Same of Figure~\ref{qu_maps} but for $I_p$. Our template 
 (left) is compared with 1.4~GHz BS76 data (right).}    
 \label{pi_maps}     
 \end{center}  
 \end{figure*}

 Our results are shown in    
 Figure~\ref{qu_maps}, where the $Q$~\&~$U$ templates at 1.4~GHz    
 are presented. Figure~\ref{pi_maps} shows a comparison between   
 the polarized intensity $I_p$ of our model and    
 the {\it $I_p$} map obtained    
 from BS76 data.    
 The comparison is performed only for $I_p$ because   
 polarization angles are strongly affected by Faraday    
 rotation at 1.4~GHz (see Section~\ref{starlight}).     
 Areas not covered   
 by our template are in grey: they correspond to the   
 Southern sky not surveyed by Reich and the region around the North   
 Celestial Pole where starlight angle data are too sparse.    
    
 The well known feature of real data, i.e. that Galactic plane and high   
 Galactic latitudes having comparable emissions, is well reproduced in our   
 template.     
      
 Our template is also able to reproduce the brightest structures in   
 BS76 data, namely:    
 \begin{enumerate}    
 \item{}the {\it Fan region} (the region situated in the Galactic plane at     
 $ 120^\circ \le l \le 150^\circ$);    
 \item{}the North Galactic Spur;    
 \end{enumerate}    
 though the latter is fainter than in real data. On the other hand, in our 
 template a feature appears in the Galactic Plane towards $l = 
 30^\circ$-$40^\circ$, which is not present in BS76. One reason might be the  
 sparse sampling of BS76 in this area, where the sampling distance is $\sim 
 4^\circ$ versus an angular resolution of $0.6^\circ$, and a source might well 
 have been missed. Another possibility could be due to Faraday depolarization 
 effects. A qualitative analysis of the whole BS76 data set (0.408-1.4~GHz) 
 suggests that at 1.4~GHz  
 only the Fan region and the North Galactic Spur are free from depolarization 
 effects. This seems to be confirmed by Junkes et al. (1990) who find the 
 polarized intensity decreasing from $l \sim 50^\circ$ towards the Galactic 
 Centre with a relevant minimum around $l \sim 30^\circ$. They argue this 
 behaviour might be due to depolarization effects: in particular, at $l \sim 
 30^\circ$, thermal material in the foreground (Scutum arm) might be responsible 
 for the observed low polarization.    
     
 The good agreement between the map obtained from BS76 data and our model    
 allows us to calibrate our template (the parameter $p \, R_{ph}$ is still free) 
 by matching the $I_p$ emission of the two maps in a well defined area. We use 
 the Fan region, the most defined and morphologically similar area in both the 
 two maps. The calibration is performed with the 820~MHz BS76 data (see Figure~2 
 in Bruscoli et al. 2002) rather than with those at 1.4~GHz, because of their 
 better sampling, providing    
 \begin{equation}   
 p \, R_{ph} = 0.9 \pm 0.09~{\rm kpc},   
 \end{equation}   
 where the error is dominated by the $\sim10$\% uncertainty on BS76 data   
 calibration.  Assuming the polarization $p$ on $7^\circ$ is in the range 
 $\sim 0.15$--0.3 (Tegmark et al. 2000) we obtain for the polarization horizon:    
 \begin{equation}    
 3 \,\mbox{kpc} < R_{ph} < 6  \,\mbox{kpc},   
 \end{equation}    
 in good agreement with present estimates (D97, Gaensler et al. 2001, Landecker 
 et al. 2002). This is a further confirmation provided by our model of the 
 relation between synchrotron $I$ and $I_p$ including a polarization horizon.    
     
 We extrapolate the $Q$ and $U$ templates at 1.4~GHz to the cosmological 
 window, and in particular to the SPOrt frequencies: 22, 32, 60, 90~GHz. We use 
 a power law with the mean synchrotron spectral index $\beta = 3.0$ found by 
 Platania et al. (1997) in the $1-19$~GHz range. We do not show the resulting 
 maps being the same at 1.4~GHz apart from the normalization. Instead, we report 
 the emission of the most important structure (the Fan region) and the mean 
 polarization level  
 \begin{equation}    
 P_{\rm rms} = \sqrt{ \left<Q^2\right> + \left<U^2\right>)}    
 \end{equation}    
 of the low emission areas (the faintest 50\% pixels) in Table~\ref{prmsTab} for 
 all the SPOrt frequencies.  
 \begin{table}    
 \begin{center}    
 \caption{Peak emission ({\it Fan region}) and $P_{\rm rms}$ of our template at
 	the four SPOrt frequencies. The $P_{\rm rms}$ is computed on the low
	emission areas (the faintest 50\% pixels).}    
 \begin{tabular}{|c|c|c|}    
 $\nu$ (GHz) & $I_p$ peak ($\mu$K) & $P_{\rm rms}$($\mu$K)\\     
\hline    
1.4 & $5 \times 10^5$ & $6.6 \times 10^4$\\ 
22 & 130 & 17\\     
32 & 43 & 5.6\\     
60 & 6.5 & 0.84\\     
90 & 1.9 & 0.25\\     
\label{prmsTab}    
\end{tabular}    
\end{center}     
\end{table}    
        
 \section{Comparisons with existing data}\label{test}      
 \subsection{Power spectra}      
       
 As a first check we compute the Angular Power Spectra (APS) of our model and 
 compare them with APS obtained from BS76 maps at 1.4 GHz (Bruscoli et al. 
 2002). 
    
 To account for the irregular sky coverage of our template we use the method 
 described by Sbarra et al. (2003) and based on $Q$ and $U$ two-point 
 correlation functions. 
   
 These are estimated directly on our maps as   
 \begin{equation}     
 \tilde{C}^X(\theta) = \Delta^X_i \Delta^X_j \hspace{1cm} X =    
 Q,U,I_p     
 \end{equation}     
   where $\Delta^X_i$ is the pixel $i$ content of map $X$, $i$ and $j$    
 identify pixel pairs at distance $\theta$.    
   
\begin{figure}     
 \begin{center}      
 \includegraphics[width=1\hsize,angle=0]{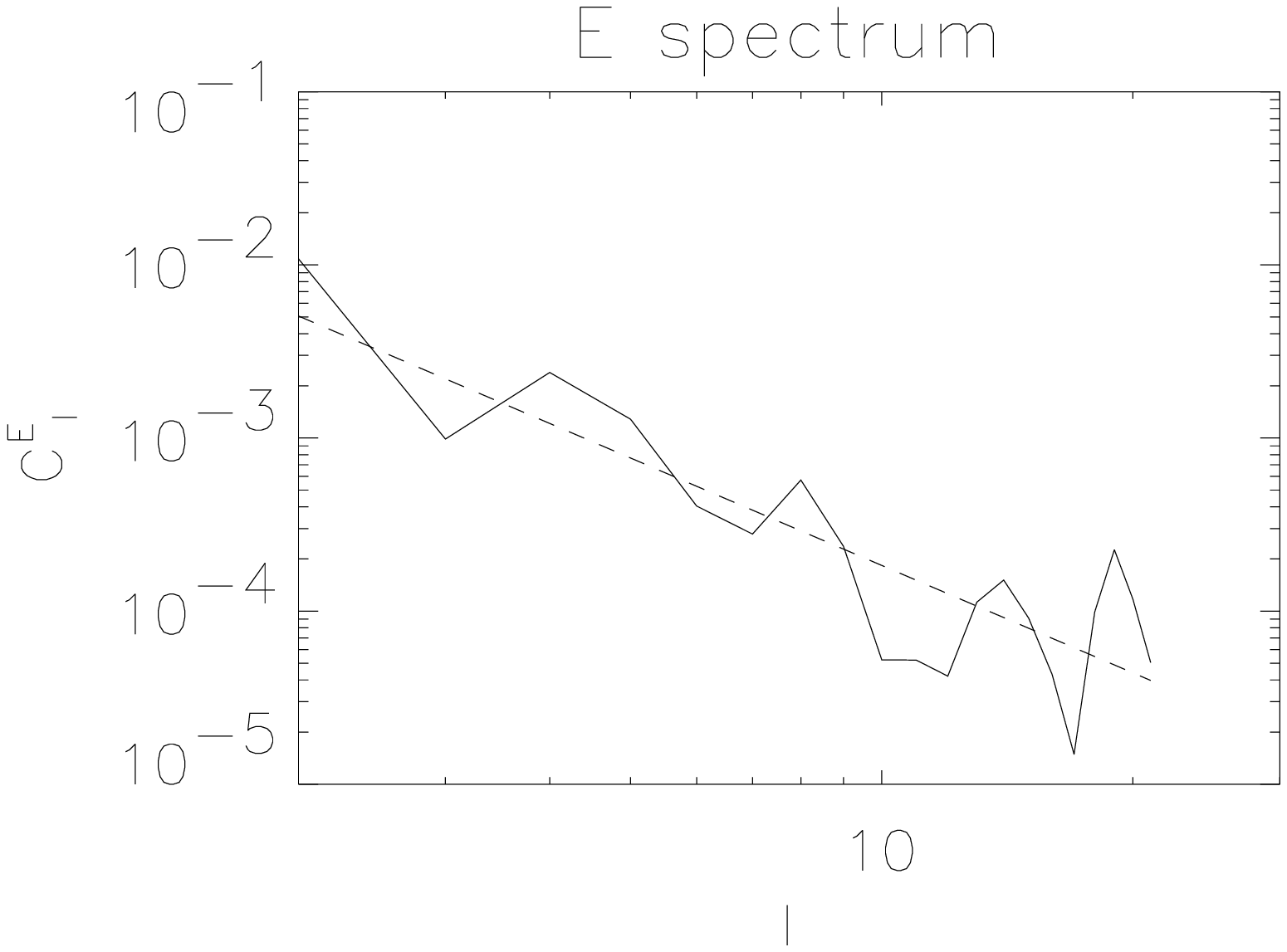}      
 \includegraphics[width=1\hsize,angle=0]{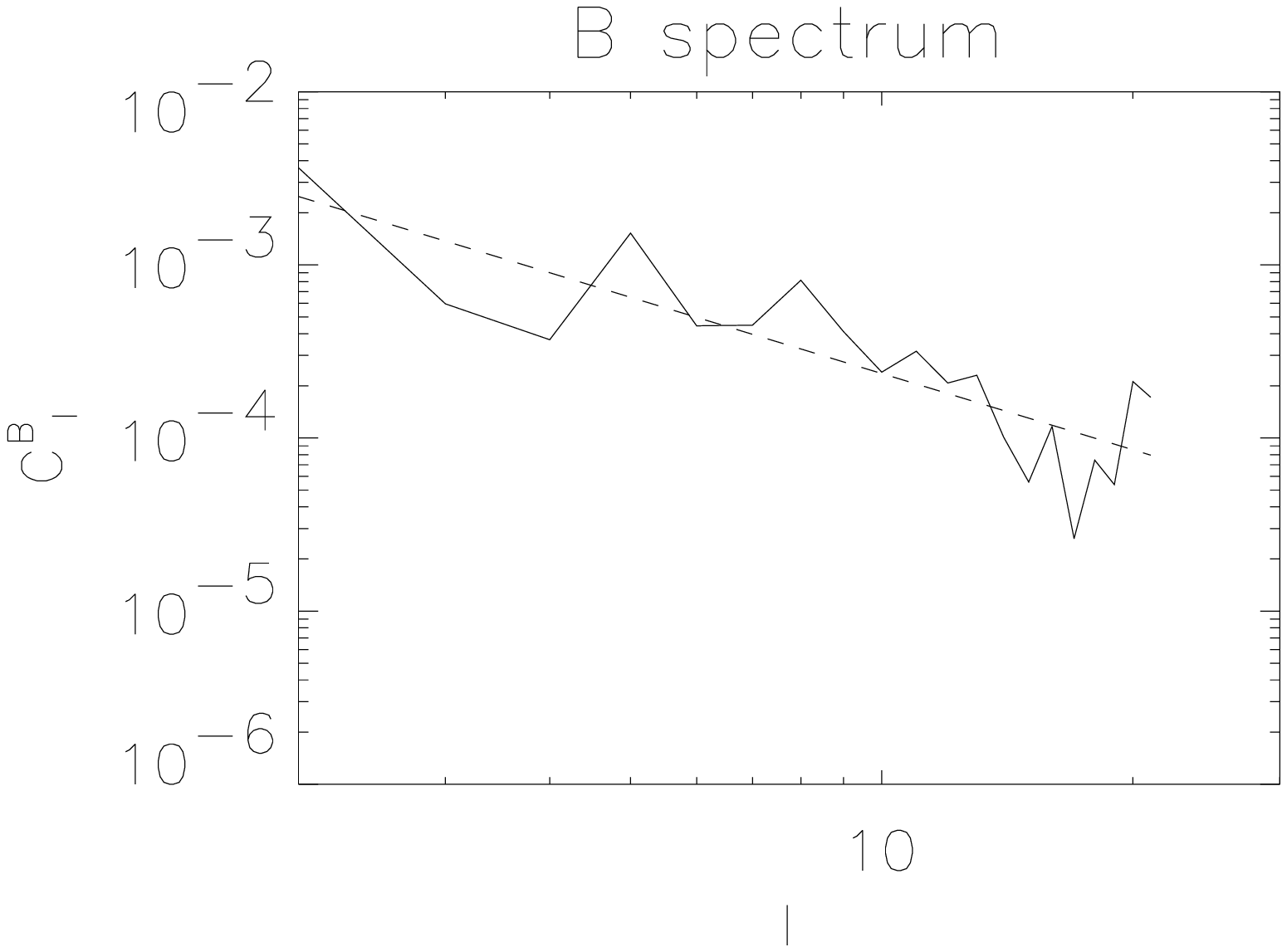}        
 \caption{$C^E$ (top) and $C^B$ (bottom) power spectra computed from our    
 synchrotron polarized emission template. Best fit curves are also shown.}      
 \label{spettro1}    
 \end{center}   
 \end{figure}

\begin{figure}      
 \begin{center}    
 \includegraphics[width=1\hsize,angle=0]{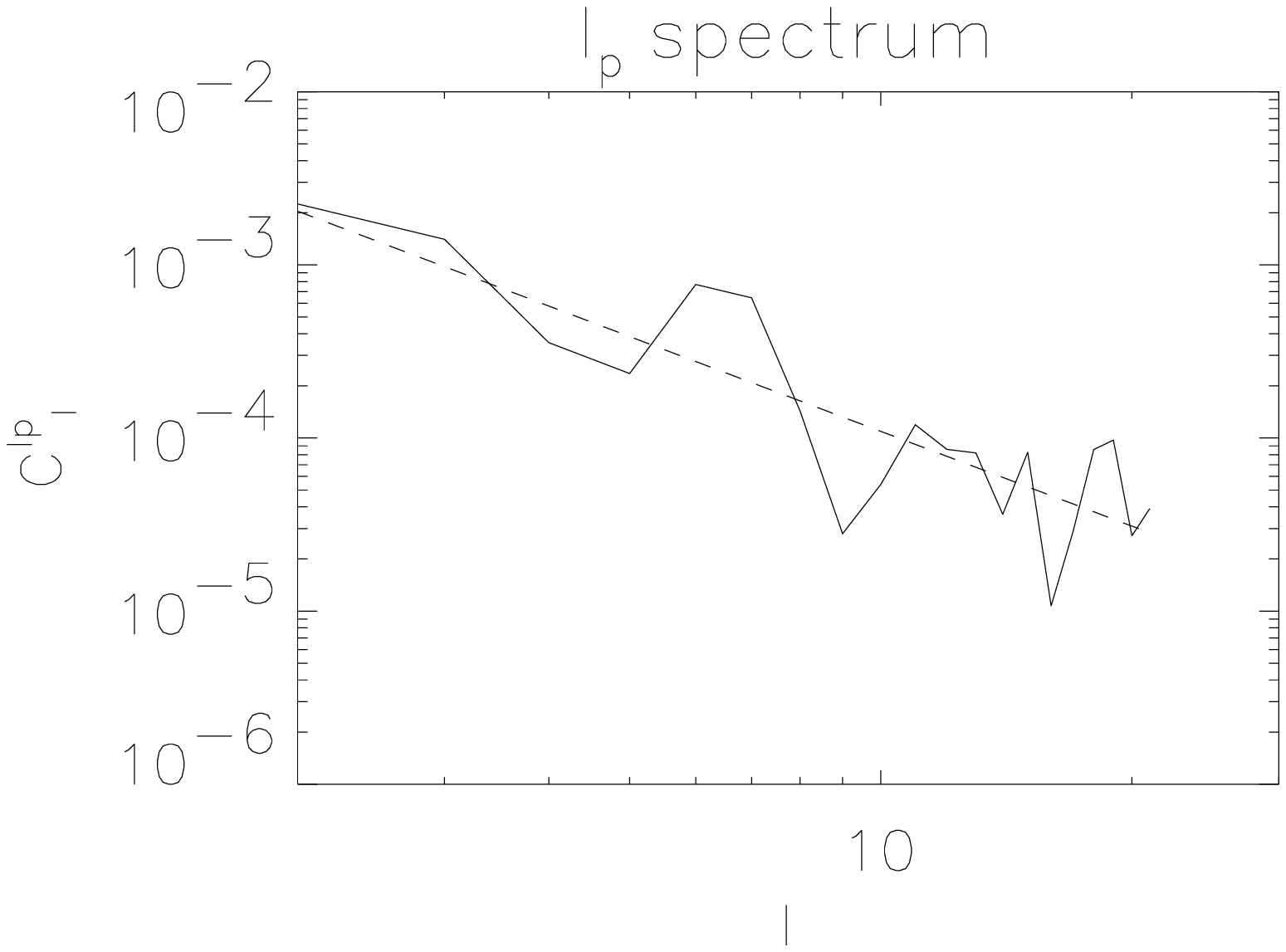}         
 \includegraphics[width=1\hsize,angle=0]{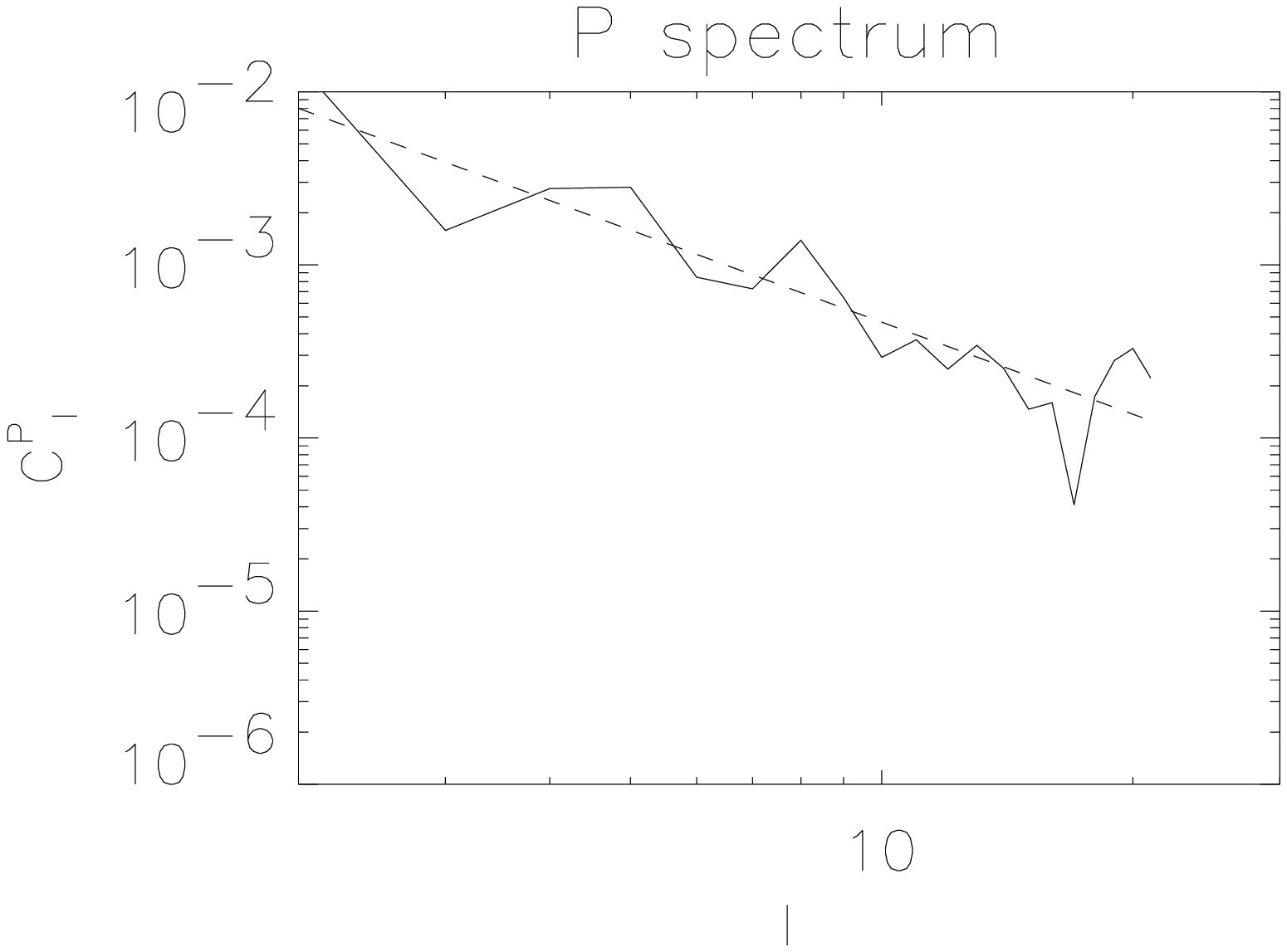}         
 \caption{The same as Figure~\ref{spettro1} but for $C^{I_p}$ (top) and    
 $C^P$ (bottom) power spectra.}      
 \label{spettro2}    
 \end{center}   
 \end{figure}

 The polarized power spectra $C^{E,B}_\ell$ and $C^{I_p}_\ell$ are obtained by 
 integration 
 \begin{eqnarray}      
  C^E_\ell    &=& W_\ell 
              \int^\pi_0 [\tilde{C}^Q(\theta)F_{1,\ell 2}(\theta) +     
                          \tilde{C}^U(\theta)F_{2,\ell 2}(\theta)]   
                          \sin(\theta)d \theta \nonumber\\     
 {C}^B_\ell   &=& W_\ell \int^\pi_0 [\tilde{C}^U(\theta)F_{1,\ell 2}(\theta) +     
                           \tilde{C}^Q(\theta)F_{2,\ell 2}(\theta)]   
                           \sin(\theta)d \theta \nonumber\\    
 {C}^{I_p}_\ell &=& {2 \pi} W_\ell \int^\pi_0 \tilde{C}^{I_p}(\theta)    
                                     P_\ell(\cos\theta) d \theta,    
 \end{eqnarray}     
 where the functions $F_{1,\ell m}$ and $F_{2,\ell m}$ are described by Zaldarriaga 
 (1998), $P_\ell$ are the Legendre polynomials, and the function      
 \begin{equation}      
           W_\ell     = e^{{\ell(\ell + 1)\sigma^2}}   
 \end{equation}   
 with $\sigma = {\rm FWHM}/\sqrt{8 \ln 2}$ accounts for beam smearing effects.   
 Finally, the total polarization spectrum $C^P_\ell$ is simply defined as   
 \begin{equation}   
     C^P_\ell = C^E_\ell + C^B_\ell.   
 \end{equation}   
 The resulting power spectra show significant    
 fluctuations also at high      
 multipoles (see Figures~\ref{spettro1} and~\ref{spettro2}).    
 We find that the errors       
 $\sigma(C_\ell)$ are significantly smaller than these    
 fluctuations, suggesting  that they are intrinsic.    
 Nevertheless, the overall behaviour of the power spectra can be   
 represented by power laws     
   \begin{equation}      
 C^Y_\ell =  A_Y \ell^{- \alpha_{Y}}   \hspace{1cm} Y = E,B,P,I_p   
 \end{equation}   
 Linear fits to the quantities $\ln C^Y_\ell$ provide the    
 values $\alpha_{Y}$ listed in Table~\ref{fit}.   
   
 Bruscoli et al. (2002) find consistent values in their analysis of    
 large portions of the BS76 maps, namely    
 $1.2 < \alpha_Y < 2$ ($1\sigma$ C.L.) in the $10 < \ell < 70$ range.    
   
 The angular behaviour of real polarized    
 synchrotron emission is thus well reproduced by our template.   
 \begin{table}    
 \begin{center}    
 \caption{Best fit APS slopes and amplitudes as obtained from our template at
 	1.4~GHz in the  $2 \le \ell \le 20$ range.} 
 \begin{tabular}{|p{0.5cm}|*{2}{l|}|}    
 	$C^Y_\ell$	& $\alpha_Y$	& $A_Y \, (K^2)$\\     
\hline    
$C^E_\ell$ & $2.1 \pm 0.3$ & $0.020 \pm 0.01$ \\     
$C^B_\ell$ & $1.6 \pm 0.2$ & $0.007 \pm 0.005$ \\     
$C^P_\ell$ & $1.9 \pm 0.2$ & $0.210 \pm 0.04$ \\     
$C^{I_p}_\ell$ & $1.9 \pm 0.3$ & $0.008 \pm 0.005$ \\     
\label{fit}    
\end{tabular}    
\end{center}     
\end{table}    

 \subsection{Free-free Emission Map}      
    
 To understand if we are subtracting    
 the right free-free contribution   
 from low frequency total intensity data, we compare our free-free      
 map with the Galactic H II region catalogue  of Kuchar \& Clarke (1997).   
    
 This is an all-sky flux compilation at $4.85$~GHz of 760~objects, representing 
 the most comprehensive H II region catalogue to date. However, a quantitative   
 comparison is not straightforward because the catalogue does not take into 
 account the diffuse component: only overall patterns can be compared.   
 \begin{figure}        
  \begin{center}  
   \includegraphics[width=0.6\hsize,angle=90]{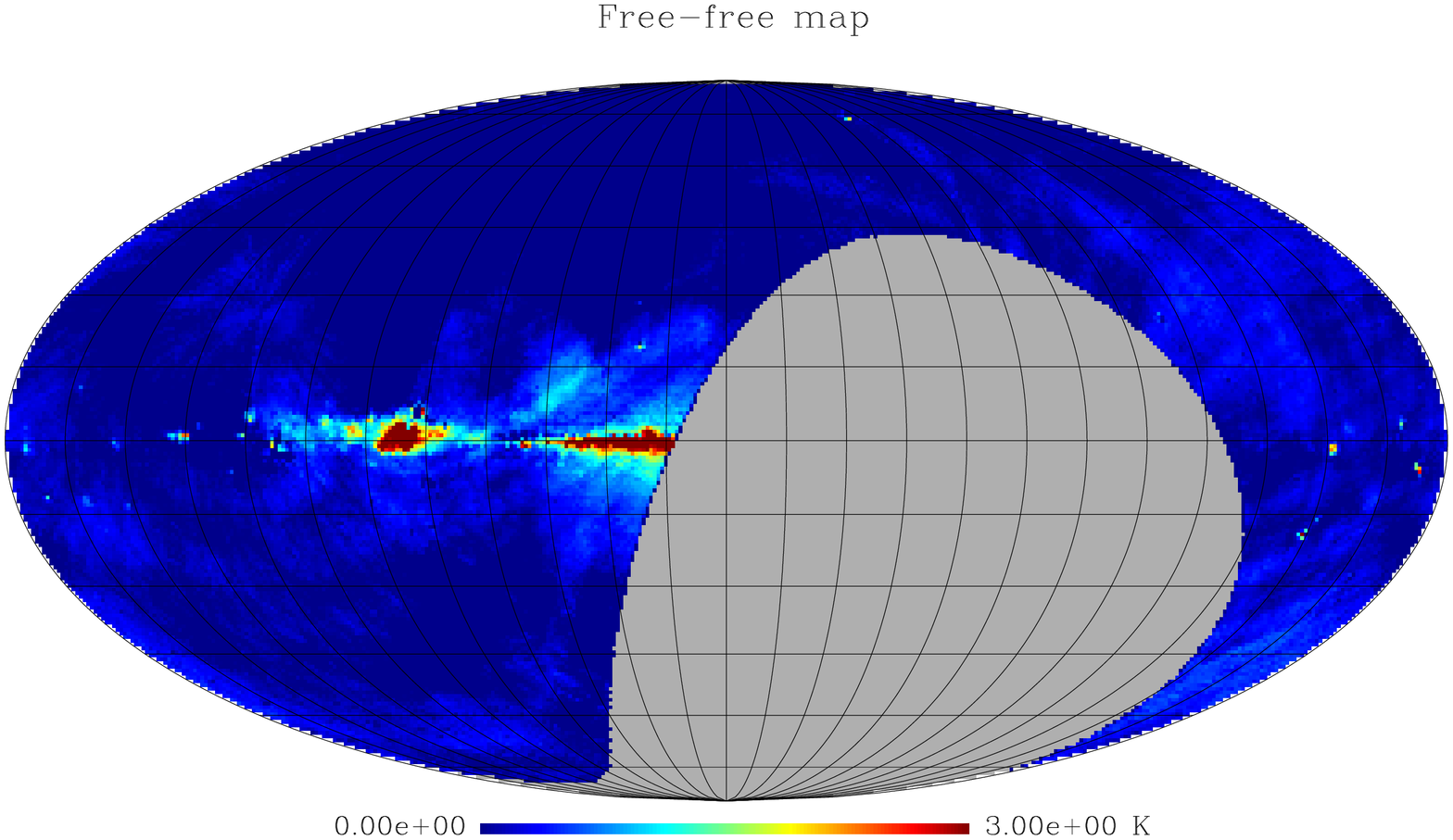}      
   \includegraphics[width=0.6\hsize,angle=90]{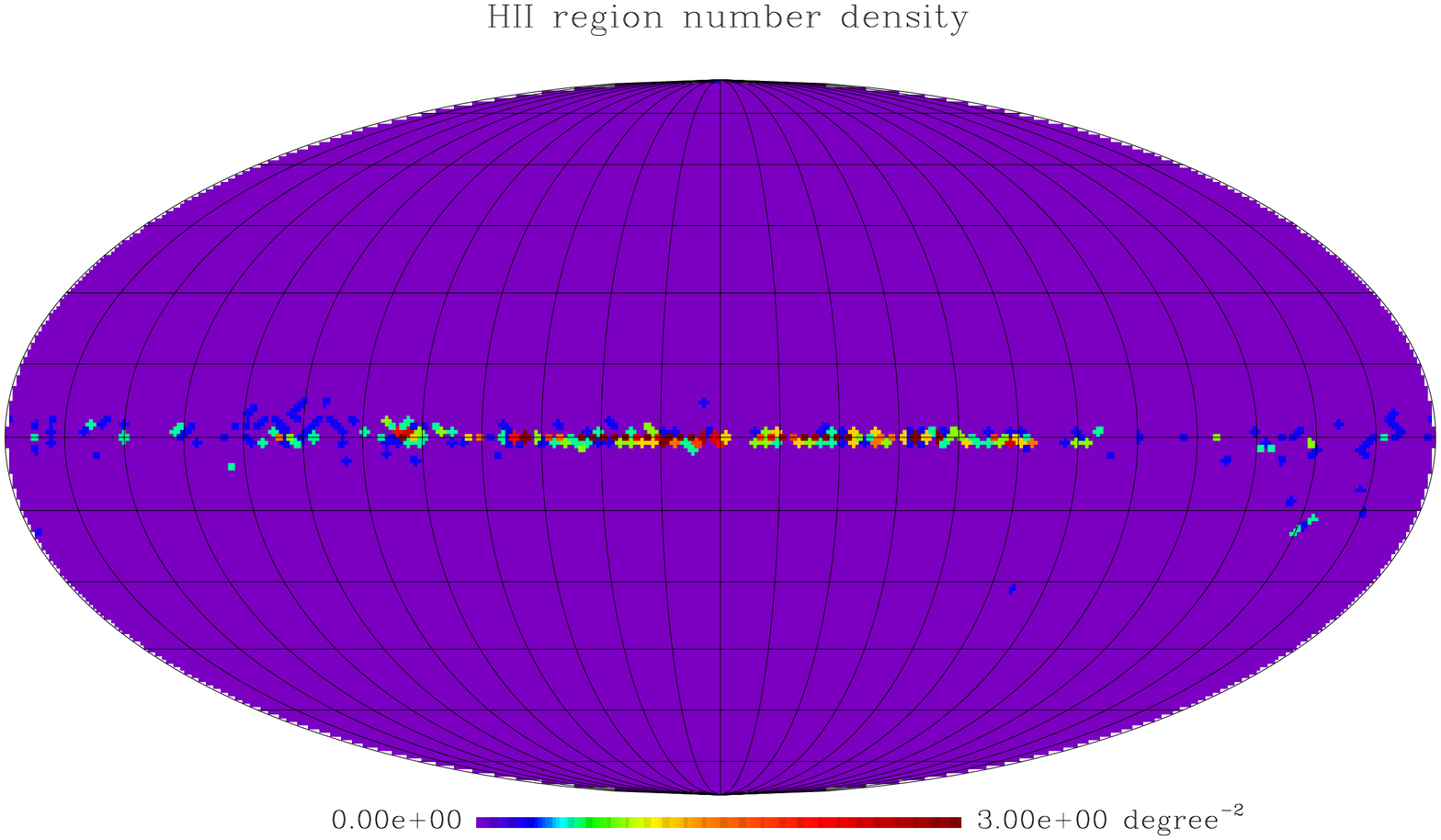}         
   \caption{Map of free-free emission at $1.4$~GHz resulting  
    from our procedure (top). Map of Galactic H II region number density  
    taken from the Kuchar \& Clarke catalogue (bottom).   
    }      
   \label{free-free_maps}     
  \end{center}  
 \end{figure}		      
\begin{figure}        
 \begin{center}        
 \includegraphics[width=1\hsize,angle=0]{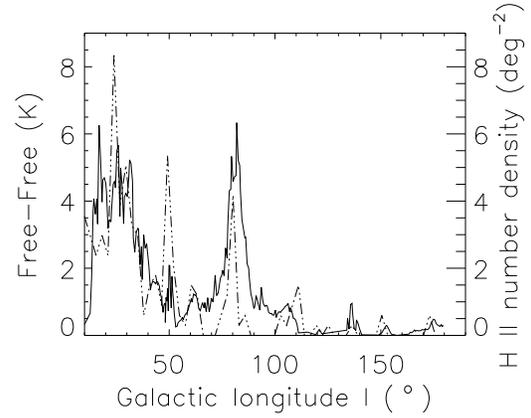}  
 \caption{Free-free emission at 1.4~GHz along the Galactic plane  
          ($b=0^\circ$) as obtained from our procedure (solid). The  
          data are smoothed on $2^\circ$ angular scale.   
           The number density of the H II region of the    
           Kuchar \& Clarke catalogue is shown for comparison (dashed).   
           }      
 \label{HIICompFig}     
 \end{center}        
\end{figure}		      
 In our free-free map (see Figure~\ref{free-free_maps}) the    
 emission is concentrated on the Galactic plane, in particular    
 towards both the Galactic Centre and the area between    
 $75^\circ < l < 90^\circ$.     
 Furthermore, it is very low at mid and high Galactic latitudes.    
 Figure~\ref{free-free_maps} also shows the    
 distribution of the Kuchar \& Clarke H II regions: they   
 too are concentrated on the Galactic plane and their number is    
 larger where our map shows the largest and strongest structures.   
 This is also suggested by Figure~\ref{HIICompFig},   
 which compares the free-free emission from our map with  
 the H II region number-density along the Galactic plane.  
 Our map is well traced by the     
 free-free emitting sources, making us confident that we are    
 subtracting thermal contributions properly.  
   
 A further test is allowed by the study of thermal     
 emission along the Galactic plane described in RR88.  
   
 A relevant contribution  (up to $15\%$ at $408$ MHz)   
 is found in the $15^\circ < l < 50^\circ$ area as well as   
 a concentration of     
 H~II regions at $l \simeq 75^\circ$ in the Cygnus,   
 where a spectral-index flattening   
 reveals the presence of a strong thermal component.   
 These areas correspond to the most evident structures in our map,   
 showing the template is also able to reproduce important free-free diffuse   
 emission sources.  
   
 Finally, a comparison of our free-free map at 1.4~GHz with    
 the Reich map at the same frequency shows that in some areas    
 on the Galactic plane the estimated free-free contribution is 50\%   
 or more of the total emission. This is an {\it a posteriori}    
 confirmation that the free-free contribution in the Haslam and Reich maps   
 is not negligible.     
           
 \subsection{The Impact of Faraday Rotation}      
       
 A check of our polarization angle map   
is performed by comparing Heiles angles with   
the data of the Parkes Southern Galactic Plane survey at 2.4~GHz (D97).   
   
However, as pointed out in Section~\ref{starlight}, Parkes data    
cannot be representative of intrinsic polarization angles    
because of their high RM values. To account for   
them we introduce a compensation by using equation~(\ref{eq12}). 
As shown by SF83, this  behaviour is correct in the Parkes area but    
 in the region around $l\sim300^\circ$, where a strong deviation is observed and 
 a {\it typical} value of RM cannot be defined.    
   
 The Parkes survey covers the Galactic plane in the      
 $l < 6^\circ$, $238^\circ < l < 360^\circ$ and $|b| < 5^\circ$ area with a      
 resolution of $\simeq 10\arcmin$ at $2.4$~GHz.     
   
 We smooth both maps on $4^\circ$ angular scale   
 to limit the impact of local RM variations. Larger scales cannot be addressed 
 since they are    
 marginally compatible with the $10^\circ$ width of the Parkes survey.   
 Furthermore, as already discussed,     
 starlight data are rotated by $90^\circ$ to match the   
 synchrotron polarization angles orientation.      
        
 Figure~\ref{map_stars} shows that    
 Heiles angles vary smoothly with the Galactic longitude.    
 Parkes data are slowly varying as well but    
 in correspondence of extended sources (D97). Here data show    
 peculiar features  (discontinuities, inversions, sudden   
 rotations) which even the smoothing procedure is not able to remove,    
 the sources extending over several degrees.    
  These regions are excluded from our comparison since the RM model   
 synthesized by equation~(\ref{eq12}) just describes the general behaviour    
  of background emission. In details,    
 following the D97 identification we exclude the regions    
  $260^\circ < l < 272^\circ$ (Vela SNR), $272^\circ < l < 285^\circ$    
 (a bright source with no total intensity counterpart),    
  $320^\circ < l < 340^\circ$ (SNR),    
 $0^\circ < l < 6^\circ$ (Galactic centre with several peculiar structures).    
     
 We divide the rest of the Parkes surveys in six patches, of     
 at least $8^\circ \times 4^\circ$, characterized by a small variation    
 of the  polarization-angle pattern.    
    
\begin{figure}      
 \begin{center}      
 \includegraphics[width=\hsize]{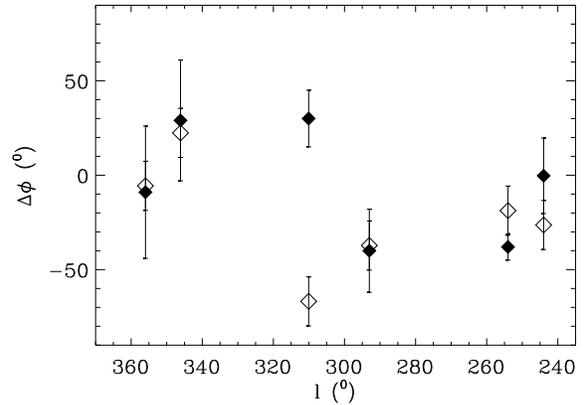}      
 \end{center}     
 \caption{Differences between polarization angles in the Parkes survey and in
 our template (filled). Faraday rotation values at 2.4~GHz as resulting from 
 equation~(\ref{eq12}) are shown for comparison (open).}       
 \label{FR_comparison}     
 \end{figure}

\begin{figure}      
 \begin{center}      
 \includegraphics[width=\hsize,angle = 0]{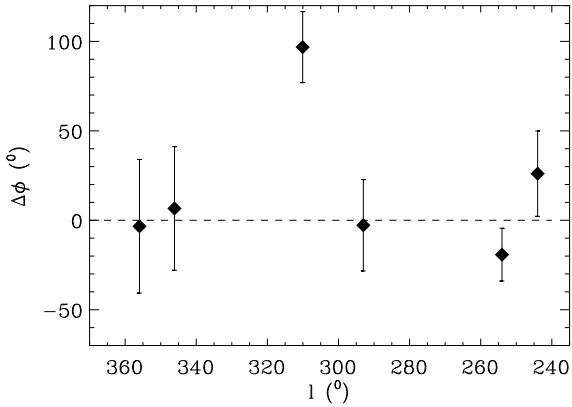}      
 \end{center}     
 \caption{Differences between the two quantities plotted in   
          Figure~\ref{FR_comparison}. The dashed line shows the  
          expected value.}      
 \label{FR_1}     
 \end{figure}

 For each selected patch we average    
 the difference between Parkes and Heiles polarization    
 angles, the latter being rotated by $90^\circ$.    
 Should Heiles angles describe the   
 magnetic field responsible for synchrotron emission,    
 these differences would match the polarization    
 angle variations induced by RM.    
 These two quantities are reported    
 in Figure~\ref{FR_comparison}, whereas their difference, expected to be    
 zero, is shown in   
 Figure~\ref{FR_1}.   
    
 The general agreement    
 between Parkes-Heiles differences and RM effects    
 confirms that Heiles data   
 provide a reliable template for the polarization angles of   
 Galactic synchrotron emission.   
   
       
 Finally, we stress that the evident disagreement    
 at approximatively $l\sim 300^\circ$ is not surprising,    
 being RM data in this area not well fitted by equation~(\ref{eq12}).    
   
 Both SF83 and D97 note that this region presents a complex    
 situation where a sudden inversion of polarization angles takes place, probably
 due to the transition from the Carena to the Centaurus arms   
 ($l \simeq 302^\circ$). Here a $180^\circ$ rotation of   
 magnetic field occurs generating large changes in RMs.    
      
 \section{Conclusions} \label{conc}    

In this paper we have presented a new approach for a template of the 
polarized Galactic synchrotron emission which, free from Faraday rotation
effects, can be better 
extrapolated to the cosmological-window frequency range (20--100~GHz). 
Differing from previous spatial models (Giardino et al. 2002, Kogut \& 
Hinshaw 2000), it is intended to provide the real spatial distribution of 
both polarized intensity and polarizationn angles. We notice that most previous 
works adopted a complementary approach based on angular frequency rather 
than real space (Tucci et al. 2000, 2002; Baccigalupi et al. 2001; Giardino 
et al. 2001; Bruscoli et al. 2002). In fact, angular spectra are commonly 
used for scale separation in the case of CMB, since they are suitable for 
cosmological parameters fitting. However, the shape of polarization angular 
spectra found at frequencies $\sim 1$ GHz, being affected by Faraday 
rotation, cannot be confidently extrapolated to the cosmological window. 
This point was raised up by Tucci et al. (2001) and Bruscoli et al. (2002), 
who noticed different behaviours in $C_l^P$ and $C_l^{I_p}$ spectra at all 
scales in the range $l\simeq 10\div 10^4$, suggesting the latter be 
less affected by Faraday rotation. In particular, the analysis of the ATCA 
Test Region at 1.4 GHz (Tucci et al. 2002) shows strong changes of slope at 
small angular scales for $C_l^X$ with $X=E,B,P$, but not for $C_l^{I_p},$ 
and this seems to be the most dramatic effect of Faraday screens. At 
present, no method is known for correcting such effects directly on angular 
spectra. The present model is intended to overcome such a problem too: 
polarization angular spectra in the cosmological window should be computed 
on the spatial template rather than simply extrapolated from the direct 
analysis of low-frequency maps. 
 
The model construction consists in three steps: 
 
\begin{description} 
\item[1)]  The synchrotron total intensity $I$ is estimated from the 
low-frequency radio surveys, cleaned from free-free emission (see Section
\ref{ingredients} and \ref{dodelson}). 
 
\item[2)]  The polarized intensity $I_p$ is estimated from $I$ by assuming that 
the Galactic polarized synchrotron emission is local because of the existence of
a polarization horizon (see Section \ref{polHorSec}). 
 
\item[3)]  The polarization angle map is built up from starlight polarization
data (see Section \ref{starlight}). 
\end{description} 
 
Step 3 is of great importance for building a pattern of Stokes parameters. 
The available RM measurements suggest in fact that the effects of Faraday
rotation on polarization angles are still too relevant in the radio-surveys at 
2.7~GHz, the highest available frequency, so that the intrinsic position 
angles cannot be estimated even with RM corrections in large portions of the 
sky. In our approach we simply overcome the problem using the starlight 
optical data (Heiles 2000). The local origin of this catalogue ($\sim $87\% 
of the stars within 2~kpc) and its frequency unaffected by Faraday rotation 
effects make it a reliable template for polarization angle of the Galactic 
synchrotron. Our analysis shows also that the sampling of the catalogue is 
compatible with the SPOrt angular resolution in all the sky but in the 
North Galactic Pole where it is too sparse. 
 
A set of checks provides the consistency of the model with existing data: 
 
\begin{itemize} 
\item  The free-free map obtained with our procedure well traces the HII 
region distribution from the Kuchar \& Clarke (1997) catalogue. This makes 
us confident about the validity of step 1. 
 
\item  The estimate provided for the distance of the polarization horizon 
(3-6~kpc) is in good agreement with the values obtained by observations, and 
the polarized intensity $I_p$ well reproduces the main structures observed 
in the BS76 data at 1.4 GHz. Both facts support the reliability of steps 1 
and 2. 
 
\item  The slopes of polarized angular power spectra $C_\ell^E$, 
$C_\ell^B$, $C_\ell^P 
$ and $C_\ell^{I_p}$ agree with those measured for large areas of the 1.4~GHz 
BS76 survey, within the large error bars declared by Bruscoli et al. 
(2002); discrepancies appear in the comparison with results at frequencies 
below 800 MHz. 
 
\item  The polarization angles of the template are in good agreement with 
those measured at 2.4~GHz (D97) and corrected for Faraday rotation 
effects in those regions where the D97 position angles show a smooth 
dependence on coordinates. 
\end{itemize} 
 
The last two items prove the validity of step 3. In this connection, we wish to
stress that a perfect agreement  
is not expected at all for angular power spectra, even in the case of the 
1.4~GHz BS76 survey. Since a conspicuous flattenig of polarization power 
spectra is attributed to Faraday rotation, we expect the power spectra 
derived from our template to be somewhat steeper than those of Bruscoli et 
al (2002). This effect is not so clear due to the large error bars and 
different sky coverages, but perhaps it is already marginally significant. 
The results in Table 2 can be compared to the weighted averages of the 1.4 
GHz angular slopes provided by Bruscoli et al. (2002), namely $\alpha 
_E=1.8\pm 0.3$ and $\alpha _B=1.4\pm 0.3.$ We note also that we are unable 
to find differences between  $\alpha _P$ and $\alpha _{PI}$ in our template. 
It is an open question, whether such slopes will be eventually found to be 
equal in synchrotron spectra for vanishing Faraday effects. 
 
In conclusion, our method results in a template of the polarized Galactic 
synchrotron emission at 1.4~GHz free from Faraday rotation effects, which can 
thus be directly extrapolated to the cosmological window frequencies. In this
range the Galactic synchrotron emission is expected to play the leading role in
the foreground contamination of CMBP data. Following Platania et al. (1998),
the extrapolation can be performed using a power law with spectral index $\beta
\sim 3.0$. As mentioned in Section~\ref{intro} this template, together with the
simulated CMBP data, provides a more reliable source map to test data processing
and foreground separation algorithms for CMBP experiments in the 20-100~GHz
range. The model has been developed so far for a  
FWHM~=~$7^{\circ }$ angular resolution, matching the needs of large scale 
CMBP experiments like SPOrt, and allows to build angular spectra only up to 
$\ell\simeq 20$. At present, the position angle data represent the major 
constraint, the Heiles data being sampled on a few degree distance. However, 
we believe that the method can be applied at subdegree scales as well, when 
a complete set of data on these scales will be available. 

\section*{Acknowledgments}     
This work has been carried out in the frame of the SPOrt experiment, a programme
funded by ASI. G.B. aknowledges a Ph.D. ASI grant. We thank the whole SPOrt    
collaboration. We acknowledge the use of HEALPix package.

 \appendix    
    
 \section[]{Distance $L(l,b)$ between the Sun and a Galactic Halo point}  
            \label{appLoS}      
       
\begin{figure}      
\begin{center}      
\includegraphics[width=0.85\hsize]{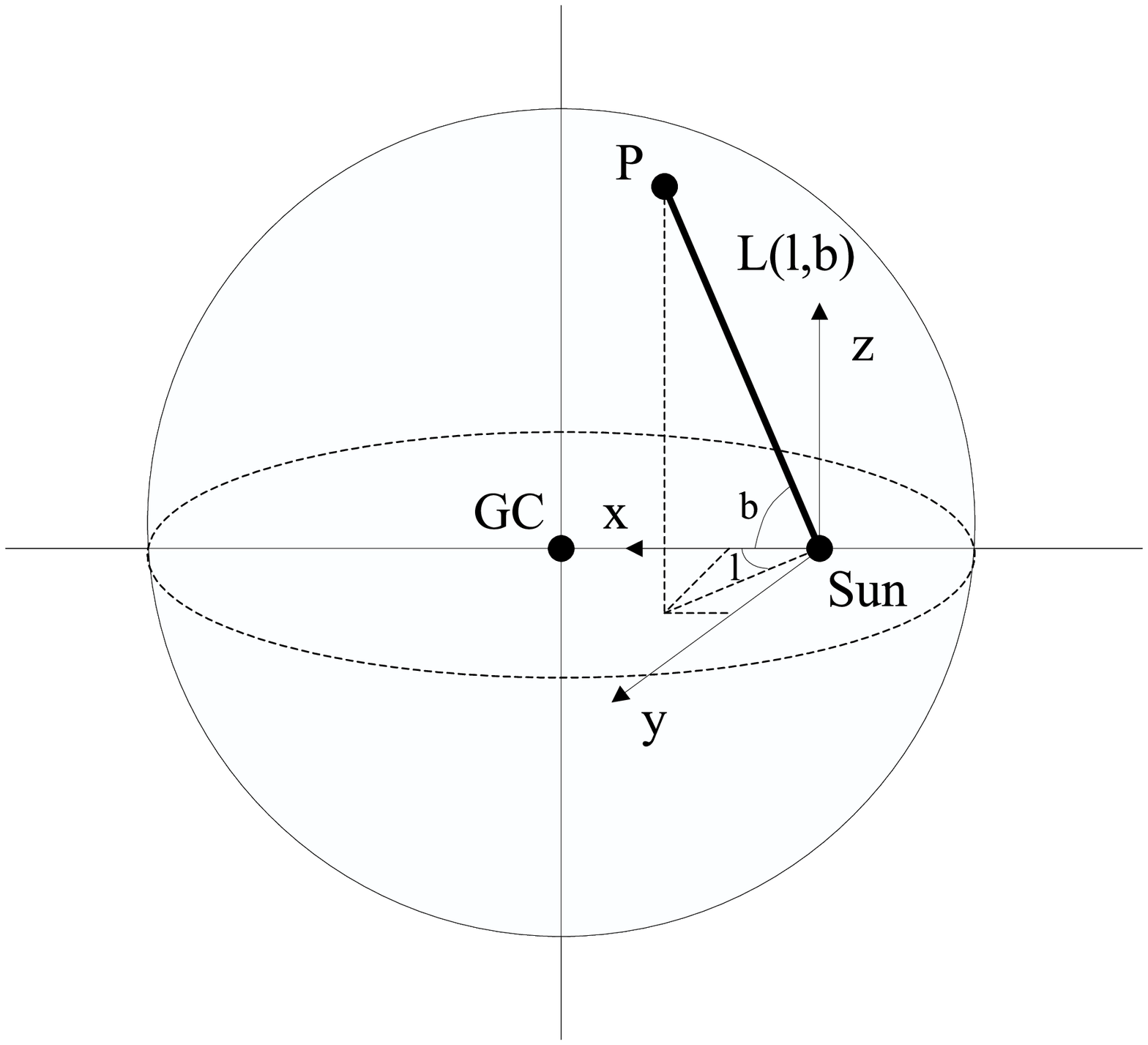}      
\end{center}     
\caption{Thickness of synchrotron emitting region $L(l,b)$ along the line of 
	 sight ($l$,~$b$) for our simple spherical model: it is the distance of 
	 the point $P$ on the halo from the Sun.    
 }      
\label{fig:pippo}     
\end{figure}

The point $P$ on the Galactic halo (radius $R = 15$~kpc) at Galactic  
coordinate ($l$,~$b$) is the intersection between the sphere representig the 
halo and the line starting from the Sun position.  
In the Sun reference frame (see Figure~\ref{fig:pippo}) this intersection  
can be expressed as:  
 \begin{equation}    
 \left\{     
 \begin{array} {ccc}    
 (x-d)^2 + y^2 + z^2 &=& R^2 \nonumber \\      
 \\  
 \displaystyle{ \frac{x}{ \cos(b) \cos(l) }} &=& \displaystyle{ \frac{y}{ \cos(b) \sin(l)}} \nonumber \\     
 \\  
 \displaystyle{ \frac{x}{ \cos(b) \cos(l) }} &=& \displaystyle{ \frac{z}{ \sin(b)}}  
 \end{array}     
 \right.     
 \end{equation}    
where $d$ is the Sun distance from the Galactic Centre (GC).   
The solution of the system provides the $P$ coordinates  
 \begin{equation} \label{eq18}     
 \left\{     
 \begin{array}{lll}      
 x \! \! \! \! &=& \! \! \! \! d \cos^2(b) \cos^2(l) \left[1 \pm \sqrt {1 + \left( \displaystyle{ \frac{R^2/d^2 - 1}      
 {\cos^2(b) \cos^2(l)} }\right)      } \right] \nonumber \\     
 y \! \! \! \! &=& \! \! \! \!  x \tan(l) \nonumber \\     
 \\ 
 z \! \! \! \! &=& \! \! \! \!  x \tan(b)/ \cos(l)      
 \end{array}     
 \right.     
 \end{equation}      
 Therefore, the distance $L(l,b)$ is given by  
 \begin{equation} \label{eq19}     
  L(l,b) = d \, \cos(b) \, \cos(l) \, \left[ 1 + \sqrt{1 + \frac{(R^2/d^2 -      
  1)}{\cos^2(b) \, \cos^2(l)}} \right]               
 \end{equation}      

\bsp     
     
\label{lastpage}     
     
\end{document}